\documentclass[11pt]{article}

\usepackage{amsfonts}
\usepackage{amssymb}
\usepackage{amsmath}
\usepackage{latexsym}

\def\beq{\begin{equation}}
\def\eeq{\end{equation}}

\def\bq{\begin{quote}}
\def\eq{\end{quote}}

\newcommand{\non}{\nonumber}
\newcommand{\be}{\begin{equation}}
\newcommand{\ee}{\end{equation}}
\newcommand{\bea}{\begin{eqnarray}}
\newcommand{\eea}{\end{eqnarray}}
\newcommand{\ba}{\begin{array}}
\newcommand{\ea}{\end{array}}
\newcommand{\al}{\alpha}
\newcommand{\bet}{\beta}
\newcommand{\pa}{\partial}
\newcommand{\ep}{\epsilon}
\newcommand{\si}{\sigma}

\newcommand{\ga}{\gamma}
\newcommand{\om}{\omega}
\newcommand{\Om}{\Omega}

\newcommand{\De}{\Delta}

\newcommand{\rar}{\rightarrow}
\newcommand{\D}{{\cal D}}

\allowdisplaybreaks
\newcounter{mycount}

\begin{document}

\begin{titlepage}
\vspace{-1mm}
\begin{flushright}
 Preprint LPT-ORSAY 01-76\\
\end{flushright}
\vspace{1mm}
\begin{center}{\bf\Large\sf Solvability of the $F_4$ Integrable System}
\end{center}

\begin{center}
{\bf Konstantin G.~Boreskov}{\normalsize
\footnote{boreskov@vxitep.itep.ru},\\ {\em Institute for
Theoretical and Experimental Physics, \\ Moscow 117259, Russia}\\
\vspace{2mm}}

{\bf Juan Carlos Lopez V.{\normalsize
\footnote{vieyra@nuclecu.unam.mx}}},
\\ {\em
Instituto de Ciencias Nucleares, UNAM, A.P. 70-543,\\ 04510 Mexico
D.F., Mexico} \\ \vspace{2mm}

{\bf Alexander V.~Turbiner{\normalsize
\footnote{turbiner@lyre.th.u-psud.fr,\
turbiner@nuclecu.unam.mx}${}^{,} $\footnote{On leave of absence
from the Institute for Theoretical and Experimental Physics, \\
\indent \hspace{5pt} Moscow 117259, Russia.}}}\\ {\em Laboratoire
de Physique Theorique, Universit\'e Paris-Sud, France and
\\Instituto de Ciencias Nucleares, UNAM, A.P. 70-543, 04510 M\'exico}
\\[2mm]
{\bf\large Abstract}
\end{center}
\small{
\begin{quote}
It is shown that the  $F_4$ rational and trigonometric integrable
systems are exactly-solvable for {\it arbitrary} values of the
coupling constants. Their spectra are found explicitly while
eigenfunctions by pure algebraic means. For both systems new
variables are introduced in which the Hamiltonian has an algebraic
form being also (block)-triangular. These variables are invariant
with respect to the Weyl group of $F_4$ root system and can be
obtained by averaging over an orbit of the Weyl group. Alternative
way of finding these variables exploiting a property of duality of
the $F_4$ model is presented. It is demonstrated that in these
variables the Hamiltonian of each model can be expressed as a
quadratic polynomial in the generators of some
infinite-dimensional Lie algebra of differential operators in a
finite-dimensional representation. Both Hamiltonians preserve the
same flag of spaces of polynomials and each subspace of the flag
coincides with the finite-dimensional representation space of this
algebra. Quasi-exactly-solvable generalization of the rational
$F_4$ model depending on two continuous and one discrete
parameters is found.
\end{quote}}
\vskip .5cm
\end{titlepage}
\vskip 2mm

\setcounter{equation}{0}

\section{Introduction}

The $F_4$ model was originally found in the Hamiltonian reduction
method in ref.\cite{Olshanetsky:1977} (see also the review
\cite{Olshanetsky:1983}). This model describes a quantum system in
four-dimensional space with Hamiltonian which is associated with
the root system $R$ of the $F_4$ algebra \cite{Olshanetsky:1983}:
\begin{align}
\label{e0.1}
 {\cal H}_{\rm F_4}^{(OP)}(z)  = & \frac{1}{2}
\sum_{i=1}^{4} \left( -\pa_{z_i}^2 + \om^2 z_i^2 \right) +
g_1 \sum_{j>i} \left\{ v(z_i-z_j) + v(z_i+z_j)\right\} \\
&+
g \sum_{i=1}^{4}v(z_i) +
g \sum_{\nu's=0,1} v\!\left( {\tfrac{z_1 + (-1)^{\nu_2}z_2+
(-1)^{\nu_3}z_3+ (-1)^{\nu_4}z_4} {2} }\right) \non\ .
\end{align}
The arguments of a potential function $v$ depend on the scalar
products $(\al,z)=\sum_{i=1}^4\al_i z_i$, and summation goes over
all positive roots $\ \al\in R_{+}=\{e_i,\, e_i\pm e_j,\,
(e_1 \pm e_2 \pm e_3 \pm e_4)/2,\ 1\leq i,j \leq 4\}$.

In the rational $F_4$ model the function $v$ takes a form
\begin{align}
\label{e0.2r}
 v(z)=\frac{1}{z^2} ~,
\end{align}
and the corresponding Hamiltonian becomes
\begin{eqnarray}
\label{e0.1r}
 {\cal H}_{\rm F_4}^{(OP,\,r)} & = & \frac{1}{2}
\sum_{i=1}^{4} \left( -\pa_{z_i}^2 + \om^2 z_i^2 \right) +
g_1\ \sum_{j>i} \left( \frac{1}{(z_i-z_j)^2} +
\frac{1}{(z_i+z_j)^2} \right)\\ &{}&+
g\  \sum_{i=1}^{4}\frac{1}{{z_i}^2} +
4g\ \sum_{ \nu's=0,1} \frac{1}{ \left[ z_1 + (-1)^{\nu_2}z_2+
(-1)^{\nu_3}z_3+ (-1)^{\nu_4}z_4 \right]^2}\nonumber\ ,
\end{eqnarray}
where $\om$ is the harmonic oscillator frequency and $g, g_1/4 >
-1/8$ are the coupling constants. The configuration space is given
by
\begin{align}
\label{e0.constr}
 -\infty < z_1 \le z_2 \le z_3 \le z_4 < \infty ~.
\end{align}
and it coincides with the Weyl chamber.

In the case of the trigonometric $F_4$ model the oscillator term
in (\ref{e0.1}) is absent, $\omega=0$, and
\begin{align}
\label{e0.2t}
 v(z;\beta) =  \frac{\bet^2}{\sin^2 \beta z} ~,
\end{align}
where $\bet$ is a parameter. Hence, the Hamiltonian takes a form
\begin{eqnarray}
\label{e0.2}
 {H_{F_4}^{(OP,\,t)}} &=& -\frac{1}{2} \sum_{i=1}^{4}
\partial_{z_i}^2 + {g_1 \bet^2}\ \sum_{j>i} \left( \frac{1}{\sin^2
\bet(z_i-z_j)} + \frac{1}{\sin^2 \bet (z_i+z_j)} \right) \\ &{}&+
{g \bet^2}\ \sum_{i=1}^{4}\frac{1}{\sin^2 \bet{z_i}} + {g\bet^2}\
\sum_{ \nu's} \frac{1}{\sin^2 \bet\frac{\left[ z_1 +
(-1)^{\nu_2}z_2+ (-1)^{\nu_3}z_3+ (-1)^{\nu_4}z_4 \right]}{2}}\ ,
\non
\end{eqnarray}
with coupling constants $g, g_1/4 > -1/8$.
The configuration space is given by the simplex
\begin{align}
\label{e0.constr_t}
 -\infty < z_1 \le z_2 \le z_3 \le z_4 < \infty
 ~,\quad z_1-z_4<\frac{\pi}{\bet}
\end{align}
and it coincides with the Weyl alcove. When $\bet$ tends to zero
the trigonometric Hamiltonian degenerates to the rational one at
$\om=0$. Both the rational and trigonometric $F_4$ models are
completely-integrable for arbitrary coupling constants $g,g_1$. If
$g=0$ the Hamiltonian (\ref{e0.1}) degenerates to the (rational or
trigonometric) $D_4$ model.

Making a change of variables in (\ref{e0.1})
\begin{align}
\label{e0.3}
 z_{1,2}=x_1 \pm x_2 \quad ,\quad  z_{3,4}=x_3 \pm x_4
\end{align}
we come to an equivalent Hamiltonian
\begin{align}
\label{e0.4} {\cal H}_{\rm F_4}(x) = &\ \frac{1}{2}\sum_{i=1}^{4}
\left(-\pa_{x_i}^2 + 4\om^2 x_i^2 \right) + 2g \sum_{j>i} \left\{
v(x_i-x_j;\beta) + v(x_i+x_j;\beta)\right\} \non\\
&+ 2g_1\!\sum_{i=1}^{4}v(2 x_i;\beta)
+ 2g_1\!\!\sum_{\nu's=0,1}\! v\left[ x_1\!+\!(-1)^{\nu_2}x_2\!+\!
(-1)^{\nu_3}x_3\!+\!(-1)^{\nu_4}x_4;\beta \right]\ .
\end{align}
Similarly to (\ref{e0.1}) this form can be associated with the
dual root system of the $F_4$ algebra $R_{+}^{\lor}=\{ 2e_i , \
e_i\pm e_j ,\ (e_1 \pm e_2 \pm e_3 \pm e_4)\}$ and we refer to the
substitution (\ref{e0.3}) as to the {\em duality} transformation.
Due to relation $v(2x;\beta)=(1/4)v(x;2\beta)$ one can see that
both in the rational and trigonometric cases this transformation
interchanges the coupling constants $g_1\rightleftarrows 2g$.
Besides that in the trigonometric case it rescales one part of the
potential (this corresponds to changing in this part $\bet \rar
2\bet$ in potential functions $v(x;\beta)$) preserving the other
part. In particular, if $g=0$ in (\ref{e0.1}) the transformation
(\ref{e0.3}) converts (\ref{e0.1}) into the Hamiltonian of $D_4$
problem.

It turns out that an analysis and formulas for the $F_4$
trigonometric problem are much simpler in $x$-coordinates than in
$z$-coordinates originally used. Similar phenomenon was observed
in $G_2$ model \cite{Rosenbaum:1998}, where among two equivalent
systems of relative coordinates (both of which were equally
suitable for the rational model) the only one led to an algebraic
form of the trigonometric Hamiltonian \footnote{The form of the
Hamiltonian is called {\it algebraic}, if exists, when the
Hamiltonian can be represented as a linear differential operator
with polynomial coefficients}.
In what follows we shall always use the form (\ref{e0.4}) of the
Hamiltonian.

In the present paper we demonstrate the exact solvability of the
rational and trigonometric $F_4$ models for general $g,g_1$. The
consideration uses a notion and a constructive criterion for exact
solvability presented in \cite{Turbiner:1994}. This notion is
based on the existence of a flag of functional spaces \footnote{A
sequence of linear spaces each one embedded into the next one, $
{\cal P}_1 \subset  {\cal P}_2 \dots \subset {\cal P}_n \subset
\dots $, forms an object called {\it flag}. A flag is called {\it
infinite flag (filtration)} if a number of these spaces is
infinite. A flag is called {\it classical} if $\mbox{dim}{\cal
P}_n=\mbox{dim} {\cal P}_{n-1}+1$.} with an {\it explicit} basis
preserved by the Hamiltonian. A  particular criterion for exact
solvability consists of checking whether the flag is related to
finite-dimensional representation spaces of a Lie algebra of
differential operators. If this criterion is fulfilled, then the
Hamiltonian of the given system can be written in terms of the
generators of this algebra which is called the {\it hidden}
algebra of the system. In \cite{Ruhl:1995} it was shown that the
eigenfunctions of the $N$-body Calogero and Sutherland models
\cite{Calogero, Sutherland:1971} form an infinite flag of linear
spaces of inhomogeneous polynomials in $(N-1)$ variables, which
coincide to finite-dimensional representation spaces of the
algebra $gl(N)$, realized by first order differential operators in
symmetric representation. The corresponding Hamiltonians were
rewritten as quadratic polynomials in the generators of the
maximal affine subalgebra of $gl(N)$, and the coupling constants
appear only in the coefficients of these polynomials. Recently, it
was shown that this statement can be extended to {\it all} $ABCD$
Olshanetsky-Perelomov rational and trigonometric integrable
systems as well as to their SUSY generalizations which turned out
to be associated to the hidden superalgebra $gl(N|N-1)$ (see Ref.
\cite{Brink:1997}). Later it was shown that for the 3-body
Calogero-Sutherland models as well as $BC_2$ models there exists a
specific additional (second) hidden algebra which was called
$g^{(2)} \subset \mbox{diff}(2,\mathbb{R})$. This algebra turned
out to be the hidden algebra of the $G_2$ rational and
trigonometric models \cite{Rosenbaum:1998,Turbiner:1998} as well.
Thus, the $A_2, BC_2, G_2$ rational and trigonometric models are
characterized by the {\it same} hidden algebra $g^{(2)}$ and their
Hamiltonians can be written as non-linear combination of the
$g^{(2)}$ generators. The flags which occurred in all
above-mentioned models were always non-classical. Therefore their
Hamiltonians were emerged in block-triangular form and the problem
of spectra was reduced to a diagonalization of each separate
block. However, a remarkable property of all above-mentioned
models holds: a certain change of variables preserving the flag
was sufficient to diagonalize all blocks simultaneously and,
finally, arrive at the pure triangular form.

In the present paper we show that the Hamiltonians of both
rational and trigonometric $F_4$ models admit algebraic form,
preserve the same infinite non-classical flag of linear spaces of
inhomogeneous polynomials and possess a hidden algebra which we
call $f^{(4)} \subset \mbox{diff}(4,\mathbb{R})$. This algebra is
an infinite-dimensional, finitely-generated algebra of
differential operators possessing a finite-dimensional invariant
subspace. It is worth to note that the $D_4$ rational and
trigonometric models, which play an important role in our
analysis, are degenerations of $BC_4$ models from one side and
$F_4$ models from another one. It implies that the $D_4$ rational
and trigonometric models possess {\it two} different hidden
algebras: $gl(5)$ and $f^{(4)}$ as a consequence of these
degenerations.

The paper is organized as follows. In Section 2 the $gl(5)$
Lie-algebraic form of the $D_4$ rational and trigonometric models
is studied. In Section 3 the rational $F_4$ model is analyzed, its
algebraic and Lie-algebraic forms are derived. In particular, a
quasi-exactly-solvable generalization of the rational $F_4$ model
is found and studied. The trigonometric $F_4$ model is
investigated in section 4. The variables providing the algebraic
and the $f^{(4)}$ Lie-algebraic forms of the Hamiltonian are found
by averaging over an orbit of the Weyl group. We also discuss an
alternative way to found such variables in connection with 'dual'
properties of the problem. Transition of the algebraic Hamiltonian
from the block-triangular to pure triangular form based on
introduction of new variables completes a demonstration of the
exact solvability of the model. A realization of the algebra
$gl(5)$ in terms of first order differential operators acting on
four-dimensional space is given in Appendix A. Appendix B is
devoted to a description of the infinite-dimensional algebra of
differential operators $f^{(4)}$, admitting finite-dimensional
representations in terms of inhomogeneous polynomials in four
variables. In Appendix C we give an explicit form of the variables
leading to the algebraic form of the $F_4$ Hamiltonian in
$z$-representation. Finally, in Appendix D the explicit formulas
for the first several eigenfunctions of the general $F_4$ model
are presented.

\setcounter{equation}{0}
\section{Algebraic and Lie-algebraic forms of the $D_4$ rational and
trigonometric models}

In this section we represent the Hamiltonians of the $D_4$
rational and trigonometric models in an algebraic form, by making
use of permutationally symmetric coordinates. The Hamiltonians can
be written in terms of the generators of the $gl(5)$ algebra
\cite{Brink:1997} and thus they have the $gl(5)$ Lie-algebraic
form.

\subsection{The rational $D_4$ model}

The Hamiltonian of the $D_4$ rational model is defined by (see
\cite{Olshanetsky:1977})
\begin{align}
\label{e1.1}
 {\cal H}^{(r)}_{D_4}(x) =
\frac{1}{2}\sum_{i=1}^{4}\left[-\pa_{x_i}^{2} + 4\om^{2}x_{i}^{2}\right] +
2g\sum_{i<j}^4 \left[ \frac{1}{(x_{i}-x_{j})^{2}} +
\frac{1}{(x_{i}+x_{j})^{2}} \right]
\end{align}
where $g = \nu(\nu - 1)/2 > -1/8$ is the coupling constant
and $\om$ is the harmonic oscillator frequency. The ground state
eigenfunction of the Hamiltonian (\ref{e1.1}) is given by
\begin{align}
\label{e1.2}
 \Psi_{0}^{(r)} \equiv \exp(-\Phi_0^{(r)}) = \left( \De_-\De_+\right)^{\nu}
\exp\left({-\om\sum_{i=1}^{4}x_{i}^{2}}\right) \ ,
\end{align}
with
\begin{align}
\label{DpDm}
 \De_{\pm} \ =\ \prod_{j<i}^4 (x_i\pm x_j) \ ,
\end{align}
and the ground state energy is
\begin{align}
E_0= 4\om (1 + 6\nu)\ .
\end{align}

As a first step towards an algebraic form of (\ref{e1.1}) let us
make a gauge rotation
\begin{align}
\label{e1.4}
 h^{\rm (r)}_{D_4} = -2 (\Psi_0^{(r)}(x))^{-1}({\cal
H}^{(r)}_{D_4}-E_0) \Psi_0^{(r)}(x) \  .
\end{align}
Then, in order to code permutation symmetry
$x_i\leftrightarrow x_j$ and reflection symmetry \ $x_i\rar -x_i$
of the Hamiltonian \footnote{ These symmetries correspond to the
group of automorphisms of the root space which in the $D_4$ case
is broader than the Weyl group.}, we take as new coordinates the
elementary symmetric (Vieta) polynomials $S_i$ of the arguments
$x_i^2$,
\begin{align}
\label{e1.5}
 s_1 = S_1(x^2) &= {x_{1}}^{2} + {x_{2}}^{2} +
{x_{3}}^{2} + {x_{4}}^{2}\ , \non\\
 s_2  = S_2(x^2)&=
{x_{1}}^{2}\,{x_{2}}^{2} + {x_{1}}^{2}\,{x_{3}}^{2}
 + {x_{1}}^{2}\,{x_{4}}^{2} + {x_{2}}^{2}\,{x_{3}}^{2} + {x_{2}}
^{2}\,{x_{4}}^{2} + {x_{3}}^{2}\,{x_{4}}^{2} \ ,\non\\
 s_3  = S_3(x^2) &= {x_{1}}^{2}\,{x_{2}}^{2}\,{x_{3}}^{2} +
{x_{1}}^{2}\, {x_{2}}^{2}\,{x_{4}}^{2} +
{x_{1}}^{2}\,{x_{3}}^{2}\,{x_{4}}^{2}
 + {x_{2}}^{2}\,{x_{3}}^{2}\,{x_{4}}^{2} \ ,
\non\\
 s_4  = S_4(x^2) &=
{x_{1}}^{2}\,{x_{2}}^{2}\,{x_{3}}^{2}\,{x_{4}}^{2}\ .
\end{align}
In the $s$ variables the Hamiltonian $h^{(r)}_{D_4}$ becomes
\begin{align}
\label{e1.6}
 h^{(r)}_{D_4} = \sum_{i,j = 1}^{4}
A_{ij}\frac{\pa}{\pa s_{i}} \frac{\pa}{\pa s_{j}}
+ \sum_{j=1}^{4} (B_{j}+C_j(\omega,\nu))\frac{\pa}{\pa s_{j}} ~,
\end{align}
where
\begin{align}
\label{e1.7}
 A_{ij} =&
\sum_{k=1}^{4} \frac{\pa s_i}{\pa x_k} \frac{\pa s_j}{\pa x_k} =
 4 \sum_{l\geq 0} (2l + 1 + j -
i)s_{i-l-1}s_{j+l} \ , \non \\[5pt]
 A_{j\,i} =&\ A_{ij} \ ,\non \\[4mm]
 B_{j}+C_j(\om,\nu) =& \sum_{k=1}^{4} \left(\frac{\pa^2
s_i}{\pa {x_k}^2} \right) - \frac{1}{2}\sum_{k=1}^{4} \frac{\pa
\Phi_0^{(r)}}{\partial x_{k}} \frac{\partial s_j}{\pa x_{k}}
\non\\[5pt] =& 2[ 1 +2\nu (4-j)](5 - j) s_{j-1} - 8\om j s_{j} \ ,
\end{align}
where $s_0 = 1$ and $s_i = 0$ at $i>4$ or $i<0$.

Since the coefficients (\ref{e1.7}) are polynomials in $s$, the
operator (\ref{e1.6}) gives the {\it algebraic} form of the
$D_{4}$ rational Hamiltonian. It is worth to note that the
coefficient matrix $A_{ij}$ makes sense of a flat space metric. An
important feature of (\ref{e1.7}) is that the coefficients
$A_{ij}$ and $B_{j}$ are the second and the first order
polynomials in $s$ coordinates, respectively. It can be shown that
it leads to two important conclusions: (i) the operator
(\ref{e1.6}) preserves the flag of spaces of polynomials
\begin{align}
\label{e1.8}
 {\cal P}_{n}^{(D_4)} \ = \ \langle s_1^{p_1} s_2^{p_2}s_3^{p_3}s_4^{p_4}
 |\ 0 \leq p_1 + p_2 + p_3+ p_4 \leq n \rangle\ ,
\end{align}
with the characteristic vector \footnote{A term
`characteristic vector' was proposed in \cite{Ruehl:1998}. It is a
vector with components which are equal to the coefficients in
front of $p_i$.}
\begin{align}
\label{e1.8p}
 \vec f \ =\ (1,1,1,1)\ ,
\end{align}
and hence the operator (\ref{e1.6}) possesses infinitely many
finite-dimensional invariant subspaces, and (ii) the operator
(\ref{e1.6}) has the $gl(5)$ Lie-algebraic form \cite{Brink:1997},
since it can be rewritten in terms of the \mbox{$gl(5)$ algebra}
generators but without raising generators $J_i^+$ (see Appendix
A). If $\nu=0$, $\om=0$, the operator $h^{(r)}_{D_4}$ becomes the
flat space Laplacian written in the \mbox{$gl(5)$ Lie-algebraic}
form.

The operator (\ref{e1.6}) with coefficients (\ref{e1.7}) appears
to be in pure triangular form with respect to the action on basis
of monomials $s_1^{p_1}s_2^{p_2}s_3^{p_3}s_4^{p_4}$. Therefore,
the spectrum of (\ref{e1.6}), $h^{\rm (r)}_{D_4} \varphi = - 2\ep
\varphi$, can be found explicitly and is equal to
\begin{align}
\label{e1.9}
  \ep_n\ =\ 4\om (p_1 + 2 p_2 + 3 p_3+ 4 p_4) \ ,
\end{align}
where $n=0,1,\ldots$ and $p_i$ are non-negative integers with a
condition $p_1 + p_2 + p_3+ p_4=n$. The spectrum does not depend
on the coupling constant $g$, it is equidistant and corresponds to
the spectrum of the harmonic oscillator. Degeneracy of the
spectrum is related to the number of partitions of an integer
number $n$ to $p_1 + p_2 + p_3+ p_4$. The spectrum of the original
rational $D_4$ Hamiltonian (\ref{e1.1}) is $E_n=E_0+\ep_n$.

It is worth to mention that the boundaries of configuration space
are determined by zeros of the ground state wave function
(\ref{e1.2}). In $s$-variables the boundary is an algebraic
surface in four variables
\begin{align}
\label{e1.10}
 \left( \De_+ \De_- \right)^2 =&
\prod_{i<j}(x_{i}^2-x_{j}^2)^{2}=
 256 s_4^3 -s_4^2(192s_3s_1+
 128s_2^2-144s_2 s_1^2+ 27s_1^4) \non\\ &
 +2s_4(72 s_3^2 s_2-3 s_3^2 s_1^2-40 s_3 s_2^2 s_1+
 9 s_3 s_2 s_1^3+8 s_2^4-2 s_2^3 s_1^2)\non\\[8pt]
 & -27 s_3^4 +2 s_1  s_3^3(9 s_2- 2 s_1^2) - s_2^2
 s_3^2(4 s_2-  s_1^2)=0 \ .
\end{align}
A simple relation between Jacobian and pre-exponential factor in
the ground state wave function (\ref{e1.10}) exists
\begin{align}
\label{e1.10jac}
 \left[\det\left( \frac{\pa  s_i}{\pa x_k} \right)\right]^2 =
 256\, \left( \De_+ \De_- \right)^2  s_4 \ .
\end{align}
Such a simplicity is of a general character because the Jacobian
for the transformation from $x$'s to the basic Weyl-invariant
polynomials is equal (up to constant factor) to the product of
linear functions vanishing on hyperplanes corresponding to roots
(see, for instance, \cite{Bourbaki}), i.e. just to $\De_+\De_-$
from (\ref{DpDm}). Since our $s$-variables (\ref{e1.5}) differ
from the $D_4$ basic invariants in taking $s_4 = (x_1 x_2 x_3
x_4)^2$ instead of $x_1 x_2 x_3 x_4$, the extra factor $s_4$
appears in the squared Jacobian.

\subsection{The trigonometric $D_4$ model}

The Hamiltonian of the $D_{4}$ trigonometric model has the form
\cite{Olshanetsky:1977}
\begin{align}
\label{e1.11}
 {\cal H}^{(t)}_{D_4} =
-\frac{1}{2}\sum_{i=1}^{4} \frac{\pa^{2}}{\pa x_{i}^{2}} +
2g\beta^2\sum_{i<j}^4\left[
\frac{1}{\sin^{2}(\beta(x_{i}-x_{j}))} +
\frac{1}{\sin^{2}(\beta(x_{i}+x_{j}))} \right]
\end{align}
where $g = \nu(\nu - 1)/2> -1/8$ and $\beta$ is a parameter. When
$\beta$ tends to zero the Hamiltonian (\ref{e1.11}) coincides with
(\ref{e1.1}) at $\om=0$.

The ground state wave function is
\cite{Olshanetsky:1983,Bernard:1995}
\begin{align}
\label{e1.12}
 \Psi_{0}^{(t)} = \left(\De_-(x,\bet)\De_+(x,\bet)\right)^{\nu}
 = \beta^{-12\nu} \prod_{i<j}\left|\sin^2 \beta x_{i} -
 \sin^2 \beta x_{j}\right|^{\nu}  \ ,
\end{align}
with
\begin{align}
\De_{\pm}(x,\bet) = \bet^{-6}\prod_{i<j}\sin \bet(x_i\pm x_j) \ ,
\end{align}
and the ground state energy equals
\begin{align}
E_0 = 28\bet^2 \nu^2 \ .
\end{align}

Using the same approach as in Section 2.1, we make a gauge
rotation of (\ref{e1.11}) with the ground state eigenfunction as a
gauge factor, $h^{(t)}_{D_4} = -2 \Psi_{0}^{-1}({\cal
H}^{(t)}_{D_4}-E_0)\Psi_{0}$. A straightforward calculation leads
to the operator
\begin{align}
\label{e1.13}
 h^{(t)}_{D_4} = \sum_{i=1}^{4} \pa_{i}^{2} +
\nu\sum_{i<j}\left[\cot\left(\beta(x_{i}-x_{j})\right)(\pa_{i} - \pa_{j})
+ \cot\left(\beta(x_{i}+x_{j})\right)(\pa_{i} + \pa_{j})\right]
\end{align}
In order to code the permutation symmetry $x_i \leftrightarrow
x_j$ and reflection symmetry $x_i \rar -x_i$, as well as the
periodicity of the Hamiltonian, we introduce new coordinates as
the elementary symmetric polynomials of the trigonometric
arguments (for definition, see eq.(\ref{e1.5}))
\begin{align}
\label{e1.14}
 \si_{k}(x) &= S_k (y_i^2) \ ,
\end{align}
\begin{align}
\label{e1.14y}
 y_i = \frac{\sin \beta x_i}{\beta} \ ,
\end{align}
which in the limit $\beta \rar 0$ coincide with (\ref{e1.5}).
In these variables the Hamiltonian $h^{(t)}_{D_4}$ becomes
\begin{align}
\label{e1.15}
 h^{(t)}_{D_4} = \sum_{i,j = 1}^{4}
A_{ij}\frac{\pa}{\pa \si_{i}} \frac{\pa}{\pa \si_{j}} +
\sum_{j=1}^{4} (B_{j}+C_j(\nu))\frac{\pa}{\pa \si_{j}}
\end{align}
where
\begin{align}
\label{e1.16}
 &A_{ij} =\ 4 \sum_{l\geq 0} (2l+1+j-i)\si_{i-l-1}\si_{j+l} \non\\
&\phantom{A_{ij} =} - 4\bet^2 [i\si_{i} \si_{j} + (i-j-2)\si_{i-1}
\si_{j+1}+(i-j-4)\si_{i-2} \si_{j+2}] \quad\text{for}\quad i \leq
j \ ,\non \\ &A_{j\,i} =\ A_{ij} \ ,\non \\[4mm] &B_{j}+C_j(\nu)
=\ 2[ 1 + 2\nu (4-j)](5 - j) \si_{j-1} - 4\bet^2 j[\si_{j} + (7-j)
\nu  \si_{j+1}] \ .
\end{align}
It is assumed that $\si_0 = 1$, and $\si_i = 0$ at $i<0$ or $i>4$.
The operator (\ref{e1.15}) with coefficients (\ref{e1.16}) gives
the {\it algebraic} form of the $D_{4}$ trigonometric Hamiltonian.
It is worth to emphasize that $A_{ij}$ makes sense of a
one-parametric flat space metric, which is reduced to (\ref{e1.7})
at $\beta=0$. Similarly to what appeared in previously discussed
$D_4$ rational case the coefficients $A_{ij}$, $B_{j}$ are
polynomials in $\si$'s of second and first degree, respectively.
Hence, the Hamiltonian $h^{(t)}_{D_4}$ can be written in terms of
$gl(5)$ generators realized as first order differential operators
\cite{Turbiner:1994} (see Appendix A), producing the $gl(5)$ {\it
Lie-algebraic} form of the $D_{4}$ trigonometric Hamiltonian.
Furthermore, it can be easily verified that the $gl(5)$
Lie-algebraic form of (\ref{e1.11}) does not contain the raising
generators $J^+_{i}$. Hence the operator (\ref{e1.15}) with
coefficients (\ref{e1.16}) preserves the same flag ${\cal
P}^{(D_4)}$ (\ref{e1.8}) of the spaces of polynomials but in $\si$
variables with the same characteristic vector (\ref{e1.8p}). If
$\nu=0$, the operator $h^{(t)}_{D_4}$ becomes the flat space
Laplacian written in $gl(5)$ Lie-algebraic form depending on a
single free parameter $\bet$
 \footnote{The operators (\ref{e1.6}) with coefficients (\ref{e1.7})
  and (\ref{e1.15}) with coefficients (\ref{e1.16})
  can be considered as two different deformations of Laplacian
  which preserve the same flag (\ref{e1.8}). An interesting
  question is about existence of other non-trivial deformations of
  Laplacian preserving the same flag.}.

The operator (\ref{e1.15}) with coefficients (\ref{e1.16}) appears
to be in pure triangular form with respect to the action on basis
of monomials $\si_1^{p_1}\si_2^{p_2}\si_3^{p_3}\si_4^{p_4}$.
Therefore the spectrum of (\ref{e1.15}) , $h^{\rm (t)}_{D_4}
\varphi = - 2\ep \varphi$, can be found explicitly
\begin{align}
\label{spe-D4t}
 \ep_n =& 2\bet^2 \left[ 5 p_1(p_1 + 2 p_2 + 2 p_3 + 2 p_4) +
 10 p_2(p_2 + 2 p_3 + 2 p_4) + 15 p_3 (p_3+ 2 p_4)
     \right.
\non\\[5pt]
 &\left. + 20 p_4^2- 3 (p_1 + 2 p_2 + 3 p_3 + 4 p_4) +
 4 \nu (3 p_1 + 5 p_2 + 6 p_3 + 6 p_4) \right] \ ,
\end{align}
where $n=0,1,\ldots$, and $p_i$ are non-negative integers with a
condition $(p_1+p_2+p_3+p_4)=n$.  The spectrum of the original
trigonometric $D_4$ Hamiltonian (\ref{e1.11}) is $E_n=E_0+\ep_n$.
This completes a demonstration of the exact solvability of the
trigonometric $D_4$ model.

The boundaries of configuration space are determined by zeros of
the ground state wave function (\ref{e1.12}). In the ${\si}$
variables the boundary is an algebraic surface in four-dimensional
space
\begin{align}\label{e1.10t}
 &\left(\De_{-}(x,\beta)\De_{+}(x,\beta)\right)^2 =
 \beta^{-24} \prod_{i<j} \left(\sin^2\beta x_{i} -
\sin^2 \beta x_{j}\right) ^2 \non\\
 &\phantom{\qquad}= \ \ 16\,\si_4\si_2^4
- 4 \left(\si_1^2\si_4+ \si_3^2\right)\si_2^3
-\left(80\,\si_4\si_1\si_3-\si_1^2\si_3^2
+128\,\si_4^2\right)\si_2^2 \non\\[5pt]
&\phantom{\qquad\quad}
+18(\si_1^3\si_3\si_4+8\si_4\si_3^2
+8\si_4^2\si_1^2+\si_3^3\si_1)\si_2
-27\si_3^4-6\si_3^2\si_1^2\si_4 \non\\[5pt]
&\phantom{\qquad\quad}
-27\si_1^4\si_4^2-4\si_1^3\si_3^3
-192\si_4^2\si_1\si_3+256\si_4^3=0
\end{align}
(cf. (\ref{e1.10})). We should emphasize that the algebraic
surface (\ref{e1.10t}) does not depend on the parameter $\bet$ and
therefore equation (\ref{e1.10t}) defines the {\it same} surface
as in equation (\ref{e1.10}) but in the space of the
$\si$-variables. It means that the configuration space of both
rational and trigonometric $D_4$ problems is the {\em same} when
is written in appropriate variables.

There exists a trigonometric generalization of (\ref{e1.10jac})
giving a connection between Jacobian and (\ref{e1.10t})
\begin{align}
\label{e1.10jac-t}
 \left[\det\left( \frac{\pa \si_i}{\pa x_k} \right)\right]^2
 = \, 256\left( \De_+(\beta) \De_-(\beta) \right)^2 \,
 \si_4\left( 1-\beta^2 \si_1 +\beta^4 \si_2 -\beta^6 \si_3+\beta^8\si_4 \right)\ .
\end{align}
In the limit $\bet \rar 0$ the equation (\ref{e1.10jac-t}) becomes
(\ref{e1.10jac}).

\setcounter{equation}{0}
\section{The rational $F_4$ model}

We shall derive in this section the algebraic and Lie-algebraic forms
of the rational $F_4$ models which lead to polynomial eigenfunctions.
This fact together with the explicit calculation of the eigenvalues,
exhibits the exact solvability of the model.

\subsection{Algebraic form}
The ground state of the rational $F_4$ model
\begin{align}
\label{e4.0}
 {\cal H}_{\rm F_4}^{(r)} = &\ \frac{1}{2}\
\sum_{i=1}^{4} \left( -\pa_{x_i}^2 + 4\om^2 x_i^2 \right) +
2g \sum_{j>i} \left( \frac{1}{(x_i-x_j)^2} + \frac{1}{(x_i+x_j)^2}
\right)\\ &+
2g_1 \sum_{i=1}^{4}\frac{1}{{x_i}^2} +
8g_1 \sum_{ \nu's=0,1} \frac{1}{ \left[ x_1 + (-1)^{\nu_2}x_2+
(-1)^{\nu_3}x_3+ (-1)^{\nu_4}x_4 \right]^2}\nonumber\ ,
\end{align}
can be written as
\begin{align}
\label{e4.1}
 & \Psi_{0}^{({\rm r})}(x) = \left(\De_-\De_+\right)^{\nu}
 \left(\De_0 \De \right)^{\mu}
\exp\left(-\om\sum_{i=1}^{4} {x_i}^2\right)\ ,
\end{align}
where $g=\nu(\nu-1)/2$, $g_1 =\mu(\mu-1)$, and
\begin{align}
 \De_{\pm} &= \prod_{j<i}^4 (x_i\pm x_j) \ , \non\\
 \De_{0} &= 2^4 \prod_{i=1}^4 x_i \ , \non \\
 \De &= \prod_{\nu's=0,1}
 \left[x_1 + (-1)^{-\nu_2}x_2 +
 (-1)^{-\nu_3}x_3+ (-1)^{-\nu_4}x_4 \right]\ ,
\end{align}
while the ground state energy is
\begin{equation}
\label{en.r}
 E_0= 4\om (1 + 6\mu+ 6\nu) \ .
\end{equation}
The transformation (\ref{e0.3}) demonstrates the `dual relation'
between two parts of the wave function
\begin{align}
\label{dd-rat}
(\De_+\De_-)^2 (x)\ =\ (\De_0\De)^2 (z) \ , \non \\
(\De_0\De)^2 (x)\ =\ (\De_+\De_-)^2 (z) \ .
\end{align}

General statement of the Hamiltonian reduction method is that any
eigenfunction of (\ref{e4.0}) can be written in a factorizable
form as
\begin{equation}
\label{e4.2}
 \Psi(x) = \Psi_0^{(r)}(x) P_{F_4}(x) \ ,
\end{equation}
where $P_{F_4}(x)$ is a polynomial in $x_i$'s. The operator
having these polynomials as eigenfunctions can be obtained by
gauge rotation of (\ref{e4.0}):
\begin{equation}
\label{e4.3}
 h_{\rm F_4}^{(r)} = -2 (\Psi_0^{(r)}(x))^{-1}
 ({\cal H}_{\rm F_4}^{(r)}-E_0) \Psi_0^{(r)}(x)\ .
\end{equation}

In order to find variables supposedly giving an algebraic form to
the original $F_4$ Hamiltonian (\ref{e4.0}) we consider as a
criterion of their choice the invariance with respect to the
symmetries of the Hamiltonian \cite{Ruhl:1995,Ruehl:1998}, i.e.
under the group of automorphisms $A$ of the $F_4$ root space (in
the $F_4$ case this group coincides with the Weyl group $W$). The
invariant polynomials of the {\it lowest} possible degrees
generate the algebra $S^W$ of $W$-invariant polynomials. These
polynomials (denoted below as $t_a^{(\Om)}$) can be found by
averaging elementary polynomials $(\alpha,x)^{2a}$ over some group
orbit (we used the orbit $\Omega$ generated by the root $e_1 +
e_2$, other orbits give algebraically related invariants):
\begin{align}
\label{orbit}
 t_a^{(\Om)}(x) = \sum_{\al\in\Om} (\al,x)^{2a}\ ,\qquad
 a=1,3,4,6 \ .
\end{align}
The powers $2a=2,6,8,12$ are the {\em degrees} of the group $W$.

These polynomials written explicitly as polynomials in $s$
(see (\ref{e1.5})) have a form
\begin{align}
t^{(\Om)}_1 =& \ s_1 \ ,\non\\ t^{(\Om)}_3 =& -12s_3+2s_2s_1+s_1^3
\ ,\non\\ t^{(\Om)}_4 =& ~ 80s_4-52s_3 s_1+\frac{20}{3}
s_2^2+s_1^4 \ ,\non\\ t^{(\Om)}_6 =& -346s_1^3 s_3+20s_2^3-720s_4
s_2+1270s_1^2 s_4+16s_2 s_1^4 +86s_2^2 s_1^2 \non\\ & -122s_1 s_2
s_3+366s_3^2+s_1^6 \ .
\end{align}

The basis of $t_a^{(\Om)}$ allows some non-linear transformations
preserving the Weyl invariance
\begin{align}
\label{e4.4t}
 t_1^{(\Om)} &\rar  t_1^{(\Om)} \ , \non \\
 t_3^{(\Om)} &\rar  t_3^{(\Om)} + a_3 (t_1^{(\Om)})^3 \ , \non\\
 t_4^{(\Om)} &\rar  t_4^{(\Om)} + a_4 (t_1^{(\Om)})^4 + b_4 t_1^{(\Om)}
 t_3^{(\Om)} \ ,\non\\
 t_6^{(\Om)} &\rar  t_6^{(\Om)} + a_6 (t_1^{(\Om)})^6+ b_6 (t_1^{(\Om)})^3
 t_3^{(\Om)} + c_6 (t_1^{(\Om)})^2 t_4^{(\Om)} + d_6 (t_3^{(\Om)})^2 \ ,
\end{align}
where $a_{3,4,6}$, $b_{4,6}$, $c_6$, $d_6$ are any numbers.
Supposedly, these transformations do not destroy an algebraic form
of the operator $h_{\rm F_4}^{(r)}$ if it appears in variables
(\ref{orbit}).

Choosing
\begin{align*}
 t_1 &= t_1^{(\Om)} \ ,\\
 t_3 &= -\frac{1}{12} t_3^{(\Om)}+\frac{1}{12}(t_1^{(\Om)})^3 \ ,\\
 t_4 &= \frac{1}{80} t_4^{(\Om)} -\frac{1}{30} t_1^{(\Om)} t_3^{(\Om)}
 + \frac{1}{48} (t_1^{(\Om)})^4 \ ,\\
 t_6 &=
 -\frac{1}{720} t_6^{(\Om)} + \frac{5}{288} (t_1^{(\Om)})^2 t_4^{(\Om)}-
 \frac{1}{27} (t_1^{(\Om)})^3 t_3^{(\Om)}+ \frac{29}{1440} (t_1^{(\Om)})^6 +
 \frac{1}{1080} (t_3^{(\Om)})^2 \ ,
\end{align*}
allows to simplify the variables to the form of polynomials of
minimal possible degrees in $s$:
\begin{align}
\label{e4.4}
  t_1 &= s_1, \non\\
  t_3 &= s_3-\frac {1}{6} s_1\, s_2 \non\\
  t_4 &= s_4-\frac {1}{4}\, s_1\, s_3+
 \frac{1}{12}\, s_2^{2} \non\\
  t_{6} &= s_4\, s_2-{\frac {1}{36}}\, s_2^{3} -
 \frac {3}{8}\, s_3^{2}+\frac {1}{8}\, s_1\,
  s_2\, s_3-\frac {3}{8}\, s_1^{2}\, s_4 \ .
\end{align}

The Hamiltonian (\ref{e4.3}) written in the coordinates
(\ref{e4.4}) has an algebraic form confirming our expectation
\begin{equation}
\label{e4.5}
 h_{\rm F_4}^{(r)}
= \, \sum_{a,\,b}A_{ab} \frac{\pa^2} {\pa
 t_a\pa t_b}\ +\, \sum_{a} \left(B_a + C_a(\mu, \nu)+
C_a (\om) \right)\frac{\pa\ } {\pa t_a} \ ,
\end{equation}
where summation goes over $a,b=1,3,4,6$ and
the coefficient functions are
\begin{alignat}{2}
\label{e4.6}
 & A_{11} = 4\, t_{1}  &\qquad
 & A_{13} = 12\, t_{3} \  ,\non\\[3pt]
 & A_{14} = 16\, t_{4}\ ,  &
 & A_{16} = 24\, t_{6}\ ,\non\\[3pt]
 &A_{33} = - { \frac {2}{3}}
\,{ t_{1}}^{2}\,{ t_{3}} + { \frac {20}{3}} \,{ t_{1}}\,{ t_{4}} \
, &
 & A_{34} = - {\frac {4}{3}}
\, t_{1}^{2}\, t_{4} + 8\, t_{6}\ , \non\\[3pt]
 & A_{36} = 16\,{ t_{4}}^{2} -
2\, t_{1}^{2}\, t_{6}\ ,  &
 & A_{44} = - 4\, t_{3}\, t_{4} -
2\, t_{1}\, t_{6}\ , \non\\[3pt]
 & A_{46} = - 4\, t_{1}\, t_{4}^{2} -
6\, t_{3}\, t_{6}\ , &
 &A_{66} = - 12\, t_{3}\, t_{4}^{2} -
 6\, t_{1}\, t_{4}\, t_{6} \ ,\non \\[3pt]
  &A_{b\,a} =\ A_{a\,b} \ ,
\end{alignat}
and
\begin{alignat}{2}
\label{e4.7}
 & B_{1} = 8\ ,&\qquad
 & B_{3} = - \, t_{1}^{2} \ ,\non \\[3pt]
 & B_{4} = - 4\, t_{3} \ ,&
 & B_{6} = - 8\, t_{1} t_{4} \ .
\end{alignat}
The coefficients $A_{ab}$ have a meaning of elements of metric
with upper indexes which corresponds to the flat space. Thus, the
operator (\ref{e4.5}) with the coefficients $A_{ab}$ and $B_{a}$
only ($C_{a}(\mu,\nu)=C_{a}(\om) =0$) is the flat space Laplace
operator.

Terms stemming from the potential part of the Hamiltonian are
proportional to $\mu$, $\nu$,
\begin{alignat}{2}
\label{e4.7_1}
 &C_{1}(\mu,\nu) =  48\,(\nu +\mu) \ , &\qquad&
  C_{3}(\mu,\nu) = -2\,(2\nu+\mu)\,  t_1^2 \ ,\non\\[3pt]
 &C_{4}(\mu,\nu) = -12\,\nu\,  t_3 \ , &&
  C_{6}(\mu,\nu) = -12\,\nu\,  t_1 t_4 \ ,
\end{alignat}
(cf. (\ref{e4.7})) and also to $\om$
\begin{alignat}{2}
\label{e4.8}
 & C_{1}(\om) = -4\,\om\, t_{1} \ , &\qquad&
   C_{3}(\om) = -12\,{ \om}\, t_{3} \ , \non\\[3pt]
 & C_{4}(\om) = -16\,{ \om}\, t_{4}\ , &&
   C_{6}(\om) = -24\,{ \om}\, t_{6} \ .
\end{alignat}
The operator (\ref{e4.5}) with the coefficients $A_{ab}$, $B_{a}$,
$C_{a}(\mu,\nu)$, and $C_{a}(\om)$ represents the {\it algebraic}
form of the $F_4$-rational model. It is easy to check that in the
$t$-coordinates the Hamiltonian $ h_{\rm F_4}^{(r)}$ preserves a
flag of polynomials ${\cal P}^{(F_4)}$, given by
\begin{equation}
\label{e4.9}
 {\cal P}_{n}^{(F_4)} \ = \
\langle  t_1^{p_1} t_3^{p_3} t_4^{p_4} t_6^{p_6}
| \ 0 \leq p_1 + 2 p_3 + 2 p_4 + 3
p_6 \leq n \rangle\ ,
\end{equation}
with the characteristic vector $\vec f\ $
\begin{equation}
\label{e4.10}
 \vec f \ =\ (1,2,2,3)\ ,
\end{equation}
(cf. (\ref{e1.8p})). This vector reminds the highest root among
short ones in the root system of the algebra $F_4$ written in the
basis of simple roots\footnote{We thank Victor Ka\^c for this
remark. The similar statement holds for $A_n$-Calogero-Sutherland
models where the flags are characterized by $\vec f \ =\
(1,1,\ldots, 1)$ \cite{Ruhl:1995} and $G_2$ models where $\vec f \
=\ (1,2)$ \cite{Rosenbaum:1998}.}. The flag (\ref{e4.9}) remains
invariant under non-linear transformations (\ref{e4.4t}). The
Hamiltonian also continues to be algebraic under these
transformations. We should mention that the first study of the
algebraic form of $F_4$ model was carried out in \cite{Ruehl:1998}
using a set of variables with the only difference from our
variables (\ref{e4.4}) in ${t}_6$
 \footnote{It corresponds to $a_6 = 0, c_6 = 3/8, b_6 = 3/32,
 d_6 = 3/8$.}.
It was found the characteristic vector $(1,2,3,5)$ in variance to
(\ref{e4.10}). However, since different ${t}_6$ variable is of the
form (\ref{e4.4t}), it preserves the same flag ${\cal P}^{(F_4)}$
(\ref{e4.9}). It implies that the flag indicated in
\cite{Ruehl:1998} is not minimal in variance to the statement made
in this article.

We were able to find one-parametric algebra of differential
operators for which there exists finite-dimensional irreducible
representation marked by an integer value of the parameter.
Furthermore, the finite-dimensional representation spaces
corresponding to different integer values of this parameter form
the infinite non-classical flag ${\cal P}^{(F_4)}$ (\ref{e4.9})
(see Appendix B). We call this algebra $f^{(4)}$. Like the algebra
$g^{(2)}$ (see \cite{Rosenbaum:1998}), the algebra $f^{(4)}$ is
infinite-dimensional but finitely-generated. The $F_4$ rational
Hamiltonian in the algebraic form (\ref{e4.5}) can be rewritten in
terms of the generators of this algebra.

The operator (\ref{e4.5}) with coefficients
(\ref{e4.6})-(\ref{e4.8}) appears to be in pure triangular form
with respect to the action on monomials
$\si_1^{p_1}\si_2^{p_2}\si_3^{p_3}\si_4^{p_4}$. One can find the
spectrum of (\ref{e4.5}), $h_{\rm F_4}^{(r)} \varphi =
-2\ep\varphi$, explicitly (cf. (\ref{e1.9}))
\begin{equation}
\label{e4.11}
  \ep_n= 2\, \om (p_1 + 3 p_2 + 4 p_3+ 6 p_4) \ ,
\end{equation}
where $n=0,1,\ldots$, and $p_i$ are non-negative integers with a
condition $p_1 + 2p_2 + 2p_3+ 3p_4=n$. The spectrum does not
depend on the coupling constants $g$, $g_1$, is equidistant and
corresponds (with different degeneracy) to the spectrum of the
harmonic oscillator as well as the rational $D_4$ model. Finally,
the energies of the original rational $F_4$ Hamiltonian
(\ref{e4.0}) are $E_n=E_0+\ep_n$.

Configuration space of the rational $F_4$ model (\ref{e4.0}) is
defined by zeros of the ground state eigenfunction, i.e. by zeros
of the pre-exponential factor in (\ref{e4.1}). These zeros also
define boundaries of Weyl chamber (see \cite{Olshanetsky:1983}).
The squared pre-exponential factor can be written as a product
of two factors. The first one
\begin{equation}
\label{e4.12}
 \big(\Delta_+\,\Delta_-\big)^{2}= - 192\, t_6^{2} + 256\, t_4^{3} \ ,
\end{equation}
corresponds to the rational $D_4$ model (\ref{e1.1}) appearing at
$g_1=0$ (cf. (\ref{e1.10})). It looks much simpler than in
\cite{Ruehl:1998}. The second factor
\begin{align}
\label{e4.12a}
 \big(\Delta_0\,\Delta\big)^{2} =& - 3072\, t_6^{2} +
4096\, t_4^{3} - 2304\, t_3^2\, t_6 - 432\, t_3^{4} +
3072\, t_1\, t_3\, t_4^{2} \non \\[3pt]
&  \mbox{} -
768\, t_1^{2}\, t_4\, t_6 + 480\, t_1^2\, t_3^2\, t_4 -
192 t_1^{3}\, t_3\, t_6 - 8 t_1^3\, t_3^3  \non \\
& +16 t_1^{4}\, t_4^{2}+ 8 t_1^{5}\, t_3\, t_4
- \frac{8}{3} t_1^{6}\, t_6 \ ,
\end{align}
corresponds to a case of the degenerate $F_4$ model, $g=0$ (as it
was noted in Introduction it is equivalent to the $D_4$ model in
dual variables). Thus, a boundary of the configuration space of
the rational $F_4$ model is confined by the algebraic surfaces
(\ref{e4.12})--(\ref{e4.12a}) of the third and seventh orders,
correspondingly\footnote{In coordinates proposed in
\cite{Ruehl:1998} these surfaces appear as of the eighth and tenth
orders, correspondingly, leading to the total algebraic surface of
the 18th order. We classify this choice of variables as {\it
non-minimal}.}, being in total given by the algebraic surface of
the tenth order.

According to general theory (see \cite{Bourbaki}) the relation
between Jacobian and  the pre-exponential factor in the
ground-state wave function (\ref{e4.1}) is straightforward:
\begin{align}
\label{e4.12jac}
 \left[\det\left( \frac{\pa t_a}{\pa x_k} \right)\right]^2
 = \frac{1}{4096} \big( \De_+ \De_- \big)^2 \big( \De_0 \De \big)^2  \ .
\end{align}

In order to find eigenfunctions one can derive recurrence
relations and then solve them out. For several first
eigenfunctions it can be done explicitly (see examples in Appendix
D). Similar to what was found previously for both the Calogero
($A_n$-rational) model \cite{Calogero, Ruhl:1995} and the $G_2$
rational model \cite{Wolfes:1974, Turbiner:1998} there exists
among eigenfunctions a family $\Phi_n ( t_1)$ which depends on a
single variable $ t_1$. This fact was already used in
\cite{Minzoni:1996} in order to construct quasi-exactly-solvable
many-body generalizations of the Calogero model (for definition of
quasi-exact-solvability, see for example \cite{Turbiner:1988}).
This family of eigenstates appears due to a fact that the
coefficients $A_{11}$, $B_1$, $C_1$ in (\ref{e4.6}) --
(\ref{e4.8}) depend on the single variable $t_1$ only. Therefore,
it is easy to verify that beside the flag ${\cal P}^{(F_4)}$
(\ref{e4.9}) the Hamiltonian $ h_{\rm F_4}^{(r)}$ preserves
another flag of polynomials ${\cal P}^{(1)}$, defined by the
spaces
\begin{align*}
 {\cal P}_{n}^{(1)}( t_1) \ = \
 \langle  t_1^{p_1} | \ 0 \leq p_1 \leq n \rangle\ .
\end{align*}
Since ${\cal P}_{n}^{(1)} \subset {\cal P}_{n}^{(F_4)}$ for any
$n$, then the flag ${\cal P}^{(1)} \subset {\cal P}^{(F_4)}$. It
leads to a degeneration of the general spectral solution for the
operator $h_{\rm F_4}^{(r)}$ to an equation
\begin{align}
\label{e4.13}
 4 t_1 \Phi''-[4\,\om\, t_1 -8(6\mu+6\nu+1)]\Phi'\
 =\ E \Phi\ ,
\end{align}
which can be solved explicitly,
\begin{align}
\label{e4.14}
 \Phi_n^{(r)} = L_n^{(12\mu+12\nu+1)}
 \left(\om  t_1\right)\ ,\ E_n= -4\,\om\, n\ ,
\end{align}
where $L_n^{(a)}$ is a Laguerre polynomial in a standard notation,
$n=0,1,2,\ldots\ $. Existence of the flag ${\cal P}^{(1)}$ (which
looks like a truly minimal flag) is a consequence of very
degenerate nature of the $F_4$ rational model. The Hamiltonian of
the $F_4$ trigonometric model does not preserve this flag.
Actually, there exist other `degenerate' flags preserved by the
$F_4$ rational Hamiltonian and thus other infinite families of
eigenstates depending on $(t_1,t_3)$ or $(t_1,t_3,t_4)$ variables
only, which can be found explicitly. They are also a consequence
of degenerate nature of the $F_4$ rational model and a reader can
easily investigate them.

\subsection{Quasi-exactly-solvable generalization
of the $F_4$ rational model}

Above-mentioned remarkable property of $h_{\rm F_4}^{(r)}$
allowing a family of eigenfunctions depending on one variable
leads to a possibility to construct a quasi-exactly-solvable (QES)
generalization (for discussion see, for example,
\cite{Turbiner:1988}) of the rational $F_4$ model. In order to do
this we will use the same trick as was used in
\cite{Minzoni:1996}. We look for a QES generalization of
(\ref{e0.1}) of a form
\begin{equation}
\label{qes.1}
 {\cal H}_{\rm F_4}^{(qes)}={\cal H}_{\rm F_4}^{(r)}+V^{(qes)}( t_1) \ .
\end{equation}
Let us make a gauge rotation (\ref{qes.1}) in the form (\ref{e4.3})
and then require that the resulting operator possesses
$ t_1$-depending family of eigenfunctions. It results in the
equation
\begin{equation}
\label{qes.2}
 h^{(qes)} \Phi \equiv
 \{4  t_1 \pa^2_{11} - [4\om  t_1 - 8(6\mu+6\nu+1)]\pa_1
 -2  V^{(qes)} \}\Phi \
 =\ -2\ep \Phi\ .
\end{equation}
where the spectral parameter $\ep$ is related to energy of the
Hamiltonian (\ref{qes.1}) through $E_n=E_0+\ep_n$ with $E_0$ given
by (\ref{en.r}).

Now one can pose a question under what condition on potential
$V^{(qes)}$, the operator $h^{(qes)}$ is Lie-algebraic. The problem
is similar to which appeared in \cite{Minzoni:1996} and the
solution is the following: make a gauge rotation of $h^{(qes)}$ in
such a way that (i) to get rid off the potential $V^{(qes)}$, and (ii)
to obtain the $sl(2)$ Lie-algebraic form
\begin{align}\label{qes.3}
h^{(qes)}_{sl(2)}( t_1) &=  t_1^{-\ga} \exp(\frac{a}{4} t_1^2)\
h^{(qes)}\  t_1^{\ga} \exp(-\frac{a}{4} t_1^2)
\non\\
&= 4 J^0_n J^- -4\om J^0_n
+2[n+4(6\mu+6\nu+1)]J^- + 4a J^+_n - 4\ga J^- \ .
\end{align}
Here
\begin{equation}
\label{qes.4}
 J^{+}_n = \tau^{2}_{1} \pa_{1} - n  t_1\ ,\ J^{0}_n =  t_1
\pa_{1} - \frac{n}{2}\ ,\ J^{-} = \pa_{1} \ ,
\end{equation}
are the generators of the $sl(2)$ algebra, and the potential
$V^{(qes)}$ should be
\begin{align}
V^{(qes)}=&\ a^2  t_1^3-2a\om t_1^2+2a[2n+1-\ga+2(6\mu+6\nu+1)]  t_1
\non\\[3pt]
&
+ \frac{2\ga[\ga+1-2(6\mu+6\nu+1)]}{ t_1}\ ,
\end{align}
where the constant terms in potential are dropped off. Finally, we
arrive in $x$-variables at the Hamiltonian ($\mathbf{x}^2 =
\sum_{i=1}^{4}{x_i}^2$)
\begin{align}
\label{qes.5}
 {\cal H}_{\rm F_4}^{(qes)} = &\ \frac{1}{2}
\sum_{i=1}^{4} \left( -\pa_{x_i}^2 + 4\om^2 x_i^2 \right) +
2g\ \sum_{j>i} \left( \frac{1}{(x_i-x_j)^2} +
\frac{1}{(x_i+x_j)^2} \right)\non \\  &+
2g_1\  \sum_{i=1}^{4}\frac{1}{{x_i}^2} +
8g_1\ \sum_{ \nu's=0,1} \frac{1}{ \left[ x_1 + (-1)^{\nu_2}x_2+
(-1)^{\nu_3}x_3+ (-1)^{\nu_4}x_4 \right]^2}\non
\\[10pt] &+
a^2 (\mathbf{x}^2)^3-2a\om(\mathbf{x}^2)^2
+2a[2n+1-\ga+2(6\mu+6\nu+1)]\mathbf{x}^2 \non \\[10pt] & +
\frac{2\ga[\ga+1-2(6\mu+6\nu+1)]}{\mathbf{x}^2}\ ,
\end{align}
where we know $(n+1)$ eigenstates explicitly. Their eigenfunctions
are of the form
\begin{equation}
\label{qes.6}
 \Psi_{0}^{({\rm r})}(x) = \left(\De_- \De_+\right)^{\nu}
 \left(\De_0 \De \right)^{\mu} (\mathbf{x}^2)^{\ga}
 P_n (\mathbf{x}^2)
 \ \exp\left[{-\om\mathbf{x}^2-
 \frac{a}{4} (\mathbf{x}^2)^2}\right] \ ,
\end{equation}
where $P_n$ is a polynomial of degree $n$. Hence we constructed
the $sl(2)$ QES deformation of the $F_4$ rational model. If in
(\ref{qes.5}) the parameter $g_1$ (and, hence, $\mu$) vanishes, we
arrive at the $sl(2)$ QES generalization of the $D_4$ rational
model. The latter differs from the $sl(5)$ QES deformation found
in \cite{Hou:1998}.

To conclude a discussion of the rational case, one can state that
the rational $F_4$ model admits the algebraic and also the $f^{(4)}$
Lie-algebraic forms. Since the Lie-algebraic form of (\ref{e4.5})
with coefficients (\ref{e4.6})-(\ref{e4.8}) contain no positive-grading
generators, an infinite family of $\nu$, $\mu$-depending polynomial
eigenfunctions of (\ref{e4.5}) occurs.

\setcounter{equation}{0}
\section{The trigonometric $F_4$ model}
\subsection{Algebraic form}

Let us consider the trigonometric $F_4$ system now.
Its Hamiltonian can be represented as
\begin{equation}
\label{e3.1}
  {\cal H}_{\rm F_4}^{(t)}(x) \ =\ -\frac{1}{2} \sum_{i=1}^{4}
 \partial_{x_i}^2 + 2g V_1(x,\beta) +
 \frac{g_1}{2} V_2(x,2\beta) \ ,
\end{equation}
where $g=\nu (\nu-1)/2$, $g_1=\mu (\mu-1)$, and
\begin{align}
\label{V1}
 V_1(x,\beta) = &\ \beta^2 \sum_{j>i} \left(
\frac{1}{\sin^2 \beta(x_i-x_j)} + \frac{1}{\sin^2 \beta(x_i+x_j)} \right)\ , \\
\label{V2}
 V_2(x,2\beta) = &\ 4\beta^2 \sum_{i=1}^{4}\frac{1}{\sin^2 2\beta{x_i}}  \non\\
+\ & 4\beta^2 \sum_{ \nu's=0,1}^4 \frac{1}{\sin^2 \beta
\left[ x_1 + (-1)^{\nu_2}x_2+ (-1)^{\nu_3}x_3+ (-1)^{\nu_4}x_4 \right]}\ .
\end{align}
The ground state of the trigonometric $F_4$ model (\ref{e3.1})--(\ref{V2}) has
the form
\begin{align}
\label{e5.1}
 \Psi_0^{(t)} (x,\beta) = \left( \De_+(x,\bet) \De_-(x,\bet) \right)^{\nu}
 \left(\Delta_0 (x,2\bet) \Delta (x,2\bet) \right)^{\mu} \ ,
\end{align}
where
\begin{align}
 \De_{\pm}(x,\beta) &=
 \beta^{-6}\prod_{j<i}\sin\bet(x_i\pm x_j) \ , \non\\
 \De_{0}(x,2\beta) &=
 \beta^{-4}\prod_{i}\sin 2\beta x_i \ , \non\\
 \De (x,2\beta) &= \beta^{-8}\prod_{\nu's=0,1}\sin \beta\left[x_1 +
 (-1)^{\nu_2}x_2 + (-1)^{\nu_3}x_3+ (-1)^{\nu_4}x_4 \right] \ .
\end{align}
Here $\De_{\pm}(x,\beta),\ \De_0(x,2\beta), \De(x,2\beta)$ are the
trigonometric analogs of the factors appearing in the rational
case (see (\ref{e4.1})). In the limit $\beta \rar 0$ they coincide
with those of the rational case, $\De_{\pm}(x,0)=\De_{\pm}(x),\
\De_0(x,0)=\De_0(x),\ \De(x,0)=\De(x)$. The ground state energy of
the Hamiltonian (\ref{e3.1}) is given by (cf. (\ref{e1.12}))
\begin{align}
\label{en.t}
 E_0 = 4\bet^2 (7 \nu^2 + 14 \mu^2 + 18 \nu \mu) \ .
\end{align}

Guided by general theory \cite{Olshanetsky:1977,Olshanetsky:1983}
and experience gained with previous studies of the
Calogero-Sutherland models \cite{Ruhl:1995, Brink:1997}, $G_2$
model \cite{Rosenbaum:1998} and the rational $F_4$ model (see
Section 3), let us check first whether there exists a family of
factorized eigenfunctions of (\ref{e3.1}) of the type
$\Psi(x)=\Psi_{0}^{(t)}(x)P_{F_4}(x)$, where the $P_{F_4}$ are
polynomials in some variables. If this is the case, there is a
chance that, in accordance with the conjectures made in
\cite{Turbiner:1994}, the trigonometric $F_4$ model also possesses
a hidden algebraic structure. In order to verify this we make
first the gauge transformation of (\ref{e3.1}) with the ground
state eigenfunction (\ref{e5.1}) as the gauge factor,
\begin{align}
\label{e5.2}
 h_{\rm F_4}^{(t)} \ =\ -2\big(\Psi_{0}^{(t)}(x)\big)^{-1}({\cal
 H}_{\rm F_4}^{(t)}-E_0) \big(\Psi_{0}^{(t)}(x)\big) \ .
\end{align}

Crucially important step is to find variables (if exist) which
lead to an algebraic form of the Hamiltonian (\ref{e5.2}). In the
rational case the relevant variables were the polynomials
generating the $S^W$ algebra of the Weyl-invariant polynomials of
$x_i$. This algebra allows the grading by means of homogeneous
polynomials. In the trigonometric case we are going to deal with
Weyl-invariant symmetric {\em trigonometric} polynomials $\tau_k$
in variables
\begin{align}
 y_i = \frac{\sin(\bet x_i )}{\bet} \ , \non
\end{align}
which can be also represented by polynomials in
\begin{align}
 \si_k(x) = S_k(y_i^2) \ , \non
\end{align}
where $S_k$ are elementary symmetric polynomials.

We impose an important requirement of a correspondence (the
`correspondence' principle, see above) that in the limit
\mbox{$\beta \rar 0$} the new variables should coincide with the
variables (\ref{e4.4}) found for the rational $F_4$ model.
Therefore, let us define the Weyl-invariant trigonometric
polynomials by averaging the elementary trigonometric polynomials
over an orbit $\Omega$ generated by the root $e_1+e_2$,
\begin{align}
\label{orbit_trig} \tau^{(\Om)}_a(x,\bet) = \sum_{\alpha\in\Omega}
\left(\frac{\sin(\bet(\al,x))}{\bet}\right)^{2a} \ , \qquad
a=1,3,4,6 \ ,
\end{align}
which reproduce the expression (\ref{orbit}) in the limit
\mbox{$\beta \rar 0$}. These polynomials $\tau^{(\Om)}_a$ in terms
of elementary symmetric polynomials $\si$'s
(\ref{e1.14}) - (\ref{e1.14y}) look as
\begin{align}
\tau^{(\Om)}_1 =&\ \si_1-\frac{2}{3}\bet^2 \si_2 \ ,\non\\
\tau^{(\Om)}_3 =& -12 \si_3+2 \si_2 \si_1+\si_1^3 + \frac{2}{3}(36
\si_4-\si_2^2-3 \si_2 \si_1^2) \non\\& +\frac{4}{3}\beta^4 \si_2^2
\si_1 -\frac{8}{27}\beta^6 \si_2^3 \ ,\non\\ \tau^{(\Om)}_4 =&\ 80
\si_4-52 \si_3 \si_1+\frac{20}{3}\si_2^2+\si_1^4
+\frac{8}{3}\beta^2 (24 \si_4 \si_1+8\si_3 \si_2-2\si_2^2
\si_1-\si_2 \si_1^3) \non\\ &+\frac{8}{27}\beta^4 (-144 \si_4
\si_2+4 \si_2^3+9 \si_2^2 \si_1^2) -\frac{32}{27}\beta^6 \si_2^3
\si_1+\frac{16}{81}\beta^8 \si_2^4 \ ,\non\\ \tau^{(\Om)}_6 =&
-720 \si_4 \si_2+1270 \si_4 \si_1^2+366 \si_3^2-122 \si_3 \si_2
\si_1-346 \si_3 \si_1^3 +20 \si_2^3 + 86 \si_2^2 \si_1^2+16 \si_2
\si_1^4  \non\\ &+\si_1^6 +\frac{4}{9}\beta^2(-864 \si_4
\si_3-2856 \si_4 \si_2 \si_1+432 \si_4 \si_1^3+24 \si_3 \si_2^2
+1182 \si_3 \si_2 \si_1^2-254 \si_2^3 \si_1 \non\\ &-84 \si_2^2
\si_1^3-9 \si_2 \si_1^5) +\frac{4}{9}\bet^4 (864 \si_4^2+952 \si_4
\si_2^2-864 \si_4 \si_2 \si_1^2-538 \si_3 \si_2^2 \si_1 +84
\si_2^4  \non\\ &+72 \si_2^3 \si_1^2+15 \si_2^2 \si_1^4)
+\frac{32}{27}\beta^6 (216 \si_4 \si_2^2 \si_1+24 \si_3 \si_2^3-10
\si_2^4 \si_1-5 \si_2^3 \si_1^3)
 \non\\
& \frac{16}{81}\beta^8 (-288 \si_4 \si_2^3+8 \si_2^5+15 \si_2^4 \si_1^2)
-\frac{64}{81}\beta^{10} \si_2^5 \si_1 +\frac{64}{729}\beta^{12} \si_2^6
\end{align}

The basis of $\tau_a^{(\Om)}$ allows some non-linear
transformations preserving the Weyl invariance, which are more
general than (\ref{e4.4t})
\begin{align}
\label{lift}
 \tau^{(\Om)}_a &\rar \tau^{(\Om)}_a + q_a (\tau^{(\Om)};\beta) \ ,\qquad a=1,3,4,6 \quad ,
\end{align}
where $q_a(\tau;\beta)$ are polynomials in $\tau$'s with
$\bet$-depending coefficients of dimensions $(2\,a)$ \footnote{The
dimension of $\tau_a$ is equal to $(2\,a)$ while for $\bet$ it is
equal to $(-1)$.}.

Following the same criterion as for rational case to have the
variables in the form of polynomials of minimal possible degrees
in $\si_a$, we get
\begin{align}
\label{e5.4}
 \tau_{1} =& \,\si_{1}-\frac{2\bet^2}{3}\si_{2}\ , \non\\
 \tau_{3} =& \,\si_3 - \frac{1}{6}\,\si_1\,\si_2-
2\beta^2(\si_{4}-\frac{1}{36}\si_{2}^2)\ , \non\\
 \tau_{4} =& \,\si_4 - {\frac{1}{4}}\,\si_1\,\si_3+
{\frac{1}{12}}\,\si_2^{2}\ , \non\\
 \tau_{6} =& \,\si_4\,\si_2 - {\frac{1 }{36}}\,\si_2^{3}-
 {\frac{3}{8}}\,\si_3^{2}+\frac{1}{8}
\,\si_1\,\si_2\,\si_3 - \frac{3}{8}\,\si_1^{2}\,\si_4\ .
\end{align}
It is worth to show the relations between the variables
$\tau^{(\Om)}_a$ and $\tau_a$:
\begin{align*}
\tau^{(\Om)}_1 =&\ \tau_1 \ , \non\\ \tau^{(\Om)}_3 =& -12
\tau_3+\tau_1^3-8\bet^2 (11 \tau_4-3 \tau_3 \tau_1) +128 \bet^4
\tau_4 \tau_1+\frac{256}{3}\bet^3 \tau_6 \ , \non\\ \tau^{(\Om)}_4
=&\ 80 \tau_4-32 \tau_3 \tau_1+\tau_1^4 +16\bet^2 (-28 \tau_4
\tau_1+3 \tau_3 \tau_1^2) +\frac{64}{3}\bet^4 (20 \tau_4
\tau_1^2+3 \tau_3^2-28 \tau_6) \non\\ &+\frac{2048}{3}\bet^6
(\tau_6 \tau_1+ \tau_4 \tau_3)+\frac{4096}{3}\bet^8 \tau_4^2 \ ,
\non\\ \tau^{(\Om)}_6 =& -720 \tau_6+1000 \tau_4 \tau_1^2-96
\tau_3 \tau_1^3+\tau_1^6+96 \tau_3^2 +8\bet^2 (904 \tau_6
\tau_1+632 \tau_4 \tau_3-392 \tau_4 \tau_1^3 \non\\& -96 \tau_3^2
\tau_1 +15 \tau_3 \tau_1^4) +64\bet^4 (-248 \tau_6 \tau_1^2+426
\tau_4^2-360 \tau_4 \tau_3 \tau_1+35 \tau_4 \tau_1^4 +15 \tau_3^2
\tau_1^2) \non\\& +\frac{512}{3}\bet^6 (-144 \tau_6 \tau_3+56
\tau_6 \tau_1^3-552 \tau_4^2 \tau_1 +126 \tau_4 \tau_3 \tau_1^2+3
\tau_3^3) +\frac{4096}{3}\bet^8 (-64 \tau_6 \tau_4 \non\\& +24
\tau_6 \tau_3 \tau_1+54 \tau_4^2 \tau_1^2 +9 \tau_4 \tau_3^2)
+\frac{65536}{3}\bet^{10} (5 \tau_6 \tau_4 \tau_1+3 \tau_4^2
\tau_3) \non\\& +\frac{131072}{9}\bet^{12}
 (\tau_6^2+6 \tau_4^3)\ ,
\end{align*}
while the inverse {\it algebraic} relations -- $\tau_a$ in terms
of $\tau^{(\Om)}_a$ -- do not exist.

Quite surprisingly, the variables $\tau_{4,\,6}$ in (\ref{e5.4})
contain no explicit dependence on $\beta$. By construction in the
limit $\bet \rar 0$ the variables $\tau_a$ coincide with the
variables $t_a$ of (\ref{e4.4}). The polynomials (\ref{e5.4}) are
algebraically independent and generate a certain algebra $S^W
(\bet)$ of Weyl-invariant trigonometric polynomials. It is quite
clear that $S^W (\bet)$ is isomorphic to the algebra $S^W$. It
simply implies that the algebra $S^W$ has $\bet$-parametric
realization (\ref{e5.4}).

Finally, the gauge-rotated operator (\ref{e5.2}) in the
coordinates (\ref{e5.4}) has an algebraic form
\begin{align}
\label{e5.5}
 h_{\rm F_4}^{(t)}
= \, \sum_{a,\,b} A_{ab} \frac{\pa^2}{\pa\tau_a\pa\tau_b}\ +\,
 \sum_{a=1} \left(B_a + C_a\right)\frac{\pa\ } {\pa\tau_a} \ ,
 \qquad a,b = 1,3,4,6 \ ,
\end{align}
where the coefficient functions are
\begin{align}
\label{e5.6}
 A_{11} &= 4\,\tau_{1}-4\beta^2{
\tau}_{1}^2-\frac{32}{3}\beta^4\tau_{3}-\frac{128}{9}\beta^6
\tau_{4}\ ,\non\\[5pt]
 A_{13} &= 12\,\tau_{3}-\frac{8}{3}\beta^2(4\tau_{1}\tau_{3}+\tau_{4})-
\frac{32}{9}\beta^4\tau_{1}\tau_{4}\ ,\non\\[5pt]
 A_{14} &=  16\,\tau_{4}
-\frac{40}{3}\beta^2\tau_{1}\tau_{4}- \frac{16}{3}\beta^4\tau_{6}
 \ ,\non \\[5pt]
 A_{16} &=   24\,\tau_{6}
- 20\beta^2\tau_{1}\tau_{6}- \frac{32}{3}\beta^4\tau_{4}^2
 \ ,\non\\[5pt]
A_{33} &=  -\frac {2}{3} \,\tau_{1}^{2}\,\tau_{3}
+\frac {20}{3} \,\tau_{1}\,\tau_{4} -
\frac{8}{9}\beta^2\,(18\tau_{3}^2 +\tau_{1}^2\,\tau_{4}
 +12\tau_{6}) \ ,
 \non\\[5pt]
A_{34} &= -\frac{4}{3} \,\tau_{1}^{2}\, \tau_{4}
+ 8\,\tau_{6}
 -\frac{4}{3}\beta^2\,(\tau_{1}\,\tau_{6}
 +12\tau_{3}\,\tau_{4})\ ,
 \non \\[5pt]
A_{36} &= 16\,\tau_{4}^{2} -
2\,\tau_{1}^{2}\,\tau_{6}
 -\frac{8}{3}\beta^2 (9\tau_{3}\,\tau_{6}
 +\tau_{1}\,\tau_{4}^2)\ ,
 \non\\[5pt]
A_{44} &= - 4\,\tau_{3}\,\tau_{4} -
2\,\tau_{1}\,\tau_{6}-24\beta^2\tau_{4}^2 \ ,\non\\[9pt]
A_{46} &= - 4\,\tau_{1}\,\tau_{4}^{2} - 6 \,\tau_{3}\,\tau_{6}-
36\beta^2 \tau_{4}\tau_{6} \ ,\non\\[9pt]
A_{66} &= - 12\tau_3\tau_4^2 - 6 \tau_{1}\tau_{4}\tau_6 -
8\beta^2(6\tau_{6}^2 +\tau_{4}^3) \ , \non \\[9pt]
  A_{b\,a} &=\ A_{a\,b} \ ,
\end{align}
and
\begin{alignat}{3}
\label{e5.7}
 &B_1 = 8-8\beta^2 \tau_1 \ , &\qquad&
B_3 = - \tau_{1}^{2}-
\frac{56}{3}\beta^2\tau_{3} -\frac{32}{9}\beta^4\tau_{4}\ ,\non\\[3pt]
 &B_4 = - 4\,\tau_{3}-\frac{88}{3}\beta^2\tau_{4}
\  , &&
 B_6 = -8 \tau_{1}\tau_{4}- 56\beta^2 \tau_6 \ .
\end{alignat}
The coefficients $A_{ab}$ have a meaning of elements of a metric
with upper indexes which corresponds to the flat space for any
value of the parameter $\beta$. Hence, the operator (\ref{e5.5})
with the coefficients $A_{ab}$ and $B_{a}$ (when $C_{a}=0$)
defines the flat space Laplacian in an {\it algebraic} form. At
$\bet\rar 0$ the expressions (\ref{e5.6})--(\ref{e5.7}) become
(\ref{e4.5})--(\ref{e4.6}). Terms stemming from the potential part
of the Hamiltonian are proportional to $\mu, \nu$
\begin{alignat}{3}
\label{e5.8}
 &C_{1} = 48(\nu+\mu) -8\beta^2(5\nu+6\mu)
\tau_1 \ , &\qquad&
 C_{3}=-2(2\nu+\mu)\tau_1^2-16\beta^2(3\nu+5\mu)\tau_3 \ , \non\\[3pt]
 &C_{4}= -12\nu\tau_3-24\beta^2 (3\nu+4\mu)\tau_4 \ ,&&
C_{6} = -12\nu\tau_1\tau_4 -48\beta^2(2\nu+3\mu)\tau_6 \ ,
\end{alignat}
(cf. (\ref{e4.7})).

Eventually, the operator (\ref{e5.5}) with the coefficients
(\ref{e5.6})--(\ref{e5.8}) presents an {\it algebraic} form of the
$F_4$ trigonometric model. It is straightforward to check that the
operator (\ref{e5.5}) with the coefficients $A_{ab}$, $B_a$, $C_a$
from (\ref{e5.6})--(\ref{e5.8}) preserves the {\it same} flag of
spaces of polynomials ${\cal P}^{(\rm F_4)}$  (\ref{e4.9}) as in
the rational case (see (\ref{e4.5}))\footnote{ Note that according
to the papers \cite{Ruehl:1998,Ruehl:1999} the rational and
trigonometric $F_4$ Hamiltonians preserve different minimal flags.
Neither of them correspond to the flag ${\cal P}^{(\rm F_4)}$
(\ref{e4.9}).}. Thus, it is evident that the $F_4$ trigonometric
Hamiltonian in the algebraic form (\ref{e5.5}) can be rewritten in
terms of the generators of the $f^{(4)}$-algebra.

\subsection{Duality}

Let us make now a short digression on a parallel between two
equivalent representations of the $F_4$ Hamiltonian related to two
dual root systems mentioned in Introduction. These representations
are connected by the `duality' transformation $\D$,
eq.(\ref{e0.3}),
\begin{align*}
 z=\D x \quad, \qquad \D^{-1}=\frac{1}{2}\D ~,
\end{align*}
and we shall refer to two forms of Hamiltonian as to $z$- and
$x$-representations, correspondingly. The functional form of the
Hamiltonian in these representations is remarkably similar (but do
not coincide) because of similarity of two root systems. For the
rational case the only change occurs in values of coupling
constants $g$ and $g_1$ while for the trigonometric case a
rescaling of the parameter $\bet$ has to be done for some terms in
the potential.

The dual transformation (\ref{e0.3}) being applied to the
Hamiltonian (\ref{e3.1}) provides the following correspondence:
\begin{align}
\label{dualrel}
 \sum_{i=1}^{4} \partial_{x_i}^2 &\rightleftarrows
 2\sum_{i=1}^{4} \partial_{z_i}^2 \ ,\non\\
 V_1(x,\beta) &\rightleftarrows V_2(z,\beta) \ ,\non\\[4pt]
 V_2(x,2\beta) &\rightleftarrows 4 V_1(z,\beta) \ .
\end{align}
It implies that after the substitution (\ref{e0.3}) we  arrive
at the equivalent Hamiltonian in the Olshanetsky--Perelomov form
\cite{Olshanetsky:1983}
\begin{align}
\label{e3.2}
 H_{F_4}^{(OP)}(z) = -\frac{1}{2} \sum_{i=1}^{4}
 \partial_{z_i}^2 + \mu (\mu -1) V_1(z,\beta) +
\frac{\nu (\nu -1)}{2} V_2(z,\beta) \ ,
\end{align}
with evident rescaling of the spectrum
\begin{align}
\label{rescal}
E \rightleftarrows E/2 \ .
\end{align}

The factors of the wave function (\ref{e5.1}) corresponding to the potentials  $V_1$
and $V_2$ also satisfy the 'duality' relations:
\begin{align}
\label{dd-trig}
 &(\De_+\De_-)(x,\beta) = (\De_0\De) (z,\beta) \ ,\non \\[3pt]
 &(\De_0\De)(x,2\beta) = (\De_+\De_-) (z,\beta) \ .
\end{align}

At $\mu =0$ (corresp. $\nu =0$) the $F_4$ Hamiltonian reduces to
the $D_4$ Hamiltonians in $x$- (corresp. $z$-) representation. The
relevant variables providing an algebraic form to the $D_4$
Hamiltonian are $\sigma_k(x,\beta)$  --- symmetric polynomials
in $y_i^2=\sin^2 (\beta x_i)/\beta^2$ (see section 2.2,\
eqs.(\ref{e1.14}) - (\ref{e1.14y})). The new variables
$\tau_a(x,\bet)$ being polynomials in $\sigma(x,\beta)$ give an
algebraic form both to $\nu$- and to $\mu$-parts of the $F_4$
Hamiltonian. This means that one can expect existence of algebraic
relations between dual variables $\tau_a(z,\beta)$ and
$\tau_a(x,\beta)$. Indeed, there exist algebraic relations which
express the variables $\tau_a(z,\beta)$ through the
$\tau_a(x,\beta)$:
\begin{align}
\label{etaz}
 \tau_a(z,\beta) = p_a (\tau (x,\beta)\,;\beta) \qquad a=1,3,4,6 \ ,
\end{align}
where $p_a(\tau\,;\beta)$ are polynomials in $\tau$'s with $\beta$-depending coefficients
of the following form
\begin{align}
\label{p_a}
 p_1(\tau;\beta) &= 2\tau_1-\frac{8}{3}\bet^2\tau_1^2+\frac{256}{3}\bet^4\tau_3 \ ,\non\\
  p_3(\tau;\beta) &= -8\tau_3-\frac{1}{3}\tau_1^3+4\bet^2\big(-16\tau_4+
  \frac{1}{18}\tau_1^4+
\frac{8}{3}\tau_1\tau_3\big)-\frac{128}{9}\beta^4\tau_1^2\tau_3+\frac{2048}{9}\beta^6\tau_3^2\
,\non\\
   p_4(\tau;\beta) &=  16\tau_4 + 4\tau_1\tau_3 +  \frac{1}{12} \tau_1^4
- 4\beta^2\big(2\tau_1\tau_4 + \frac{1}{3} \tau_1^2\tau_3 \big) +
 \frac{16}{3}\beta^4\big( 2\tau_6 + \tau_3^2 \big) \ ,\non \\
  p_6(\tau;\beta) &= - 64\tau_6 - 24\tau_3^2 - 8\tau_1^2\tau_4 -
 2\tau_1^3\tau_3 - \frac{1}{36} \,\tau_1^6 \non\\ &
 - 4\beta^2\big(8\tau_3\tau_4 - 16\tau_1\tau_6 - 4\tau_1
\tau_3^2 - \tau_1^3\tau_4 - \frac{1}{6} \tau_1^4\,\tau_3\big) \non \\&
 - 16\beta^4\big(6\tau_4^2 + 2\tau_1\tau_3\tau_4
+ \frac{1}{3} \tau_1^2\tau_6 + \frac{1}{3} \tau_1^2\tau_3^2\big) +
\frac{32}{3}\bet^6 \big(\tau_3\tau_6 +  \frac{1}{3} \tau_3^3 \big)
\ .
\end{align}
The inverse relations for $\tau(x,\bet)$'s as functions of
$\tau(z,\bet)$'s are not algebraic. However it is possible to
express algebraically $\tau(x,\bet)$'s through the
$\tau(z,\bet/2)$'s with the use of eqs.(\ref{etaz}):
\begin{align}\label{etax}
 \tau_a(x,\bet) = p_a(\tau(\D^{-1} x,\bet);\bet) =
p_a(\tau(z/2,\bet);\bet) \ .
\end{align}

The relations (\ref{etax}) suggest another way of finding
variables in which the $F_4$ model acquires an algebraic form ---
one has to look for algebraic relations between $\si_k=S_k(\sin^2
(\bet x_i)/\bet^2)$ and
$\check{\si}_k=S_k(\sin^2 (\bet z_i/2)/\bet^2)$:
\begin{align}
\label{dual}
 f_a(\si;\beta)=g_a(\check{\si};\bet) \ .
\end{align}
Polynomial functions $f_a$ and $g_a$ can be used as new variables
which give an algebraic form to the Hamiltonian in $x$ and $z$
representations, correspondingly. Historically we used namely
relations (\ref{dual}) for finding relevant variables
(\ref{e5.4}): $$\tau_a\equiv f_a (\si;\bet)\ .$$ Explicit formulas
for $g_a (\check{\si};\beta)$ are given in Appendix C. One can see
that the functions $f_a$ have much simpler form than the functions
$g_a$ what gives a preference to the $x$-representation.

\subsection{Wave functions and energies of the trigonometric $F_4$ model}

Configuration space of the trigonometric $F_4$ model is defined by
zeros of the ground state eigenfunction (\ref{e5.1}), i.e. of the expression
$(\De_+\De_-\De_0\,\De)^{2}$. This expression can be
written as a product of two factors, (i):
\begin{align}
\label{DpDmt}
 \big(\De_+\,\De_-\big)^2\ =\ 256 \tau_4^3 - 192 \tau_6^2 \ ,
\end{align}
which corresponds to the trigonometric $D_4$ model appearing at
$g_1=0$, and (ii): $(\De_0\,\De)^{2}$,  which corresponds to the
degenerate $(g=0)$ $F_4$ model (cf. (\ref{e4.12}), (\ref{e4.12a})).
In order to find the second factor we use the
second relation (\ref{dd-trig}). It tells that this factor is
equal to $\De_+^2 \,\De_-^2$ written in $z$-variables and, thus,
it can be expressed through $\tau_{4,\,6}(z,\beta)$,
\begin{align}
(\De \De_0)^2 = 256\tau_4(z,\beta)^3 - 192\tau_6(z,\beta)^2 \ .
\end{align}
The use of eqs.(\ref{etaz}), (\ref{p_a}) gives
\begin{align}
\label{DD0}
 \De_0^{2}\,&\De^{2} =\big( - \frac{8}{3} \tau_1^6\tau_6 +
 8\tau_1^5\tau_3\tau_4 + 16\tau_1^4\tau_4^2 - 8\tau_1^3\tau_3^3 -
 192\,\tau_1^3\tau_3\tau_6  + 480\tau_1^2\tau_3^2\tau_4
\non\\
&
- 768\tau_1^2\tau_4\tau_6 + 3072 \tau_1\tau_3\tau_4^2 -
432\tau_3^4  - 2304\tau_3^2\tau_6 + 4096\tau_4^3 - 3072\tau_6^2
\big)
\non\\
&
+\frac{8}{3} \beta^2\big(\tau_1^7\tau_6 - 3\tau_1^6\tau_3\tau_4
- 6\tau_1^5\tau_4^2 + 3\tau_1^4\tau_3^3 + 96\tau_1^4\tau_3\tau_6 -
246\tau_1^3 \tau_3^2 \tau_4
\non\\[3pt]
&
  + 288\tau_1^3 \tau_4\tau_6 -1120\tau_1^2 \tau_3 \tau_4^2+ 144\tau_1 \tau_3^4 -
1536\tau_1\tau_4^3 + 960\tau_ 1 \tau_3^2\tau_6 + 1536\tau_1
\tau_6^2
\non\\
&
- 288\tau_3^3\tau_4 - 768\tau_3\tau_4\tau_6\big)
+\frac{16}{3}\beta^4\big( 246\,\tau_1^3 \tau_3 \tau_4^2
+768\,\tau_1 \tau_3 \tau_4 \tau_6
+ 312\,\tau_1 \tau_3^3 \tau_4
\non\\
&
- 324\,\tau_1^2\tau_3^2\tau_6 -24\tau_3^2 \tau_4^2 - 68\tau_1^4 \tau_4\tau_6
+ 30\tau_1^4 \tau_3^2 \tau_4-192\,\tau_4^2\tau_6 - 11\tau_1^5 \tau_3 \tau_6
\non\\
&
-672\tau_1^2 \tau_6^2 +360\tau_1^2 \tau_4^3 - 24\tau_ 1^2 \tau_3^4 \big)
- \frac{32}{3}\beta^6 \big(\tau_1^5 \tau_4\tau_6 - 3 \tau_1^4\tau_3
\tau_4^2 - 40 \tau_1^3\tau_3^2 \tau_6
\non\\
&
 - 6\tau_1^3\tau_4^3 -
48\tau_1^3\tau_6^2+ 96\tau_1^2\tau_3^3\tau_4+ 8\tau_1^2
\tau_3\tau_4\tau_6+ 312\tau_1 \tau_3^2\tau_4^{2} - 96\tau_1\tau_4^2\tau_6
- 48 \tau_3^5\non\\
&
- 272 \tau_3^3\tau_6 +432\tau_3\tau_4^3 - 384\tau_3\tau_6^2
\big)
+\frac{64}{9}\beta^8 \big(\tau_1^4\tau_6^2 +24\tau_1^3\tau_3\tau_4\tau_6
- 72\tau_1^2\tau_3^2\tau_4^2 \non\\
&
+ 180\tau_1^2 \tau_4^2\tau_6 -
648\tau_1\tau_3\tau_4^3 - 144\tau_1\tau_3^3\tau_6 - 384\tau_1
\tau_3 \tau_6^2
 + 288 \tau_3^4\tau_4
+1056\tau_3^2\tau_4\tau_6 \non\\
& - 972\tau_4^4 +768\tau_4\tau_6^2\big)
 -\frac{1024}{9}\beta^{10} \big(\tau_1^2\tau_3\tau_6^2
+6\tau_1\tau_3^2\tau_4\tau_6 + 24\tau_1\tau_4 \tau_6^2
-18\tau_3^3\tau_4^2 \non\\
&
 - 54\tau_3\tau_4^2\tau_6\big)
 +\frac{4096}{27}\beta^{12} \,\big(3\tau_3^{2}
+8\tau_6\big)\tau_6^{2} \ ,
\end{align}
(cf. (\ref{e4.12a})), which corresponds to the degenerate $F_4$
model at $g=0$. A boundary of the configuration space of the
trigonometric $F_4$ model is confined by the algebraic surfaces
(\ref{DpDmt}), (\ref{DD0}) of the third and eighth orders,
correspondingly. It is quite surprising that in the rational case
the corresponding surface defined by (\ref{e4.12a}) is of the
seventh order while in (\ref{DD0}) there exist the only two terms
of the eighth order: $(8/3)\beta^2( \tau_1^7\tau_6
-3\tau_1^6\tau_3\tau_4)$.

It is remarkable that in the $F_4$ trigonometric case the relation between
the Jacobian and the ground-state wave function has the same simple form
as in the rational case:
\begin{align}
\label{jac_trig}
 \left[\det\left( \frac{\pa \tau_a}{\pa x_k} \right)\right]^2
 = \frac{1}{4096} \big( \De_+ \De_- \big)^2 \big( \De_0 \De \big)^2  \ .
\end{align}
(cf.(\ref{e4.12jac})).

Let us now proceed to finding the spectrum of the Hamiltonian
(\ref{e5.5}). It is easy to see that the operator (\ref{e5.5})
with the coefficients (\ref{e5.6})--(\ref{e5.8}) has a block
triangular form in the $\tau$ variables unlike pure triangular
form which is needed in order to find eigenvalues. In general, in
order to reduce this operator to pure triangular form it is
necessary to diagonalize each block separately, doing it one by
one. Surprisingly, in our particular problem it can be performed
in full generality just by introducing unique set of new
variables~(!)
\begin{align}
\label{e5.9}
&\rho_1 = \tau_1 \ ,\non\\[5pt]
&\rho_3 = \tau_3 - \frac{1}{8}\bet^{-2}\tau_1^2 \ ,\non\\[5pt]
&\rho_4 = \tau_4 - \frac {3}{16}\bet^{-4}\tau_1^2 \ ,\non\\[5pt]
&\rho_6 = \tau_6 - \frac{3}{4}\bet^{-2}\tau_1 \tau_4
 + \frac{3}{64}\bet^{-6}\tau_1^3 \ ,
\end{align}
having the same dimension as in(\ref{e5.4}). It is worth to note
that this substitution becomes singular at $\bet = 0$, reflecting
the non-existence of bound states for the rational $D_4$ and $F_4$
models in absence of the harmonic oscillator term in potential.
This coordinate transformation is of the type (\ref{lift}) and
hence leaves the flag (\ref{e4.9}) invariant. Thus, we arrive at a
conclusion that among Weyl-invariant coordinate systems (of
minimal dimension) there exists unique one which leads to pure
triangular form of the Hamiltonian with respect to a basis of
monomials. We were unable to see a relation with variables
introduced in \cite{Ruehl:1999, Khastgir:00} in framework of a
general study of trigonometric Hamiltonians based on root system
which should guarantee triangular form.

In the coordinates (\ref{e5.9}) the Hamiltonian (\ref{e5.2}) takes
the form
\begin{align}
\label{e5.10}
 h_{F4}^{(t)} =
&
 \Big(4\rho_1 - 8\beta^2\rho_1^2 - \frac{32}{3}\beta^4\rho_3 -
 \frac{128}{9}\beta^6\rho_4 \Big)\pa^2_{\rho_1^2} + 8\Big[3\rho_3 -
 2\beta^2\Big(\frac {1}{3} \rho_4 + \rho_3\rho_1\Big)\Big]
 \pa^2_{\rho_1 \rho_3} \non\\
&+ \Big( 3\beta^{-4}\rho_1^2+32\rho_4
  + 8\rho_3\rho_1 - 24\beta^2\rho_4\rho_1 -\frac{32}{3}\beta^4\rho_6
 \Big) \pa^2_{\rho_1 \rho_4} +
 2\Big[3\bet^{-2}\rho_1\rho_4 \non\\
&+24\rho_6-4\beta^2\Big(\rho_1\rho_6
 - 2\rho_3\rho_4\Big) \Big] \pa^2_{\rho_1 \rho_6}
 -\Big[6\bet^{-2}\rho_1\rho_3 +16\bet^2\Big(\rho_3^2 +
 \frac{2}{3}\rho_6\Big)\Big]\pa^2_{\rho_3^2} \non\\
&
 + \Big(- 9\bet^{-4}\rho_1\rho_3+16\rho_6 - 32\bet^2\rho_3\rho_4 +
 6\bet^{-2}\rho_1\rho_4 \Big) \pa^2_{\rho_3 \rho_4} \non\\
&
 + \Big[-6\beta^{-2}\Big(3\rho_3\rho_4 -4\rho_1\rho_6\Big)+
 36\rho_4^2 -48\beta^2\rho_3\rho_6 \Big] \pa^2_{\rho_3 \rho_6} \non\\
&
 + \Big[-\frac{27}{16}\bet^{-8}\rho_1^3 -\frac{3}{4}\bet^{-4}\rho_1
 \Big(3\rho_1\rho_3 - 16\rho_4\Big)+2\rho_1\rho_6 - 4\rho_3\rho_4 -
 24\bet^2\rho_4^2\Big] \pa^2_{\rho_4^2} \non\\
&
 -\Big[+\frac{27}{4}\beta^{-6}\rho_1^2\rho_4+ 18\bet^{-4}\rho_1\rho_6+
 3\bet^{-2}\rho_4\Big(3\rho_1\rho_3+ 8\rho_4\Big)+12\rho_3\rho_6 \non\\
&
+ 64\bet^2\rho_4\rho_6  \Big] \pa^2_{\rho_4 \rho_6}
 - \Big[\frac{27}{4}\beta^{-4}\rho_1\rho_4^2-9\beta^{-2}\rho_6
 \Big(\rho_1\rho_3-4\rho_4\Big) +18\rho_3\rho_4^2 +48\beta^2\rho_6^2
 \Big] \pa^2_{\rho_6^2} \non\\
&
 + 8\Big[\Big(1 + 6\nu + 6\mu\Big) -\bet^2\rho_1 \Big(1 - 6\mu -
 5\nu\Big)\Big]\pa_{\rho_1} -
\Big[ 3 \beta^{-2} \Big(1+4\nu +4\mu\Big)\rho_1\non\\
&
+16\bet^2 \Big(1+3\nu + 5\mu\Big)\rho_3 \Big]\pa_{\rho_3}
-\Big[ \frac{9}{2}\beta^{-4}\Big(1 +4\nu +4\mu \Big)\rho_1+
 12\nu \rho_3 \non\\ &
+ 24\bet^2\Big(1+ 3\nu+4\mu\Big)\rho_4 \Big]
 \pa_{\rho_4} -\Big[ \frac{27}{8}\bet^{-6}\rho_1^2
- 9\bet^{-4}\nu\rho_1\rho_3+ 6\bet^{-2}\Big(5+6\nu  \non\\
 & +6\mu
 \Big)\rho_4 + 48\bet^2\Big(1+2\nu +3\mu\Big)\rho_6
\Big]\pa_{\rho_6} \ ,
\end{align}
and it seems simpler than the operator (\ref{e5.5}) with the
coefficients (\ref{e5.6})--(\ref{e5.8}). It is easy to check that
the operator (\ref{e5.10}) is indeed a triangular operator. It is
evident that this operator can be rewritten in terms of the
$f^{(4)}$-generators like it was for the operator (\ref{e5.5}).

Using the representation (\ref{e5.10}) the energy levels of the
Hamiltonian, $h_{\rm F_4}^{(t)} \varphi = -2\ep\varphi$, can be
found explicitly and are given by
\begin{align}
\label{e5.18}
 \ep_{n} =
&
 4\bet^2 [p_1 (p_1 + 2 p_3 + 3 p_4 + 4 p_6) +
 2p_3(p_3+2p_4+3p_6)+p_4(3p_4+8p_6) \non\\
&
 +6p_6^2 + \nu (5 p_1   +  6 p_3  + 9 p_4  + 12 p_6) +
 2\mu (3 p_1 + 5 p_3  + 6 p_4 + 9 p_6)]
\end{align}
where $n=0,1,\ldots$, and quantum numbers $p_a$ are non-negative
integers with a condition $p_1 + 2p_3 + 2p_4+ 3p_6=n$. The
spectrum of the original trigonometric $F_4$ Hamiltonian
(\ref{e3.1}) is $E_n=E_0+\ep_n$ (cf. (\ref{spe-D4t})).

The explicit expressions for the first several eigenfunctions of
(\ref{e5.10}) in $\rho$-variables are presented in Appendix D. It
is worth to mention that the equations describing the boundary of
the configuration space remain algebraic in $\rho$-variables and
are given by
\begin{align}
\label{e5.C}
 \big(\De_+\,\De_-\big)^2 =& -36\beta^{-6}\rho_1^3\rho_6
 +36\beta^{-4}\rho_1^2\rho_4^2-288\beta^{-2}\rho_1\rho_4\rho_6+256 \rho_4^3 - 192 \rho_6^2 \ ,
\end{align}
and
\begin{align}
\label{e5.D}
 \big(\Delta\Delta_0\big)^2 =&
- 72\beta^{-6}
\,{\rho_1}^{3}\Big(3\,{\rho_3}^2+8\,\rho_{{6}}\Big)
+ 36\beta^{-4}\, {\rho_1}^2\Big(24\,{\rho_1}^2\rho_6+9\,
{\rho_1}^2{\rho_3}^2+16\,{\rho_4}^2\Big) \non\\
&
-144\beta^{-2}\,\rho_1 \Big( 12\,{\rho_3}^2\rho_4+8\,\rho_
{{3}}\rho_1\rho_6+32\,\rho_6\rho_4+3\,{\rho_3}^{3}
\rho_1+6\,{\rho_1}^2{\rho_4}^2 \Big) \non\\
&
- \Big(3072\,{\rho_6}^2 + 4096\,{\rho_4}^{3} +
864\,{\rho_1}^{3}{\rho_3}^{3} + 2880\,\rho_4{\rho_1}^2{\rho_3}^2 +
2304\,\rho_6{\rho_1}^{3}\rho_3  \non\\
&
 +768\,\rho_3{\rho_4}^2\rho_1+ 7680\,\rho_4{\rho_1}^2\rho_6 -
2304\,{\rho_3}^2\rho_6 - 432\,{\rho_3}^4 \Big) \non\\
&
-192{\beta}^2\,\Big( 9\,{\rho_1}^2\rho_3{\rho_4}^2-32\,\rho_{{1
}}{\rho_6}^2-3\,\rho_1{\rho_3}^4-20\,\rho_1{\rho_{{3}}}^2\rho_6+
6\,\rho_4{\rho_3}^{3} \non\\
&
+36\,{\rho_4}^{3}\rho_1+16\,\rho_3\rho_4\rho_6 \Big)
+64{\beta}^4\,\Big( 60\,\rho_1\rho_4{\rho_3}^{3}+12\,{\rho_1}^{2
}{\rho_3}^2\rho_6-2\,{\rho_4}^2{\rho_3}^2 \non\\
&
+9\,{\rho_1}^2{\rho_3}^4-16\,{\rho_4}^2\rho_6-32\,{\rho_1}^2{\rho_6}^2+
160\,\rho_1\rho_3\rho_4
\rho_6 \Big)
+\frac{256}{3}{\beta}^6\,\Big( -54\,{\rho_4}^{3}\rho_3 \non\\
&
+108\,{\rho_4}^2\rho_
{{6}}\rho_1+27\,{\rho_3}^2{\rho_4}^2\rho_1+34\,
\rho_6{\rho_3}^{3}+48\,{\rho_6}^2\rho_3+6\,{\rho_{{3 }}}^{5}\Big) \non\\
&
-\frac{256}{3}{\beta}^{8}\,\Big( -88\,{\rho_3}^2\rho_4\rho_6 -
64\,\rho_4{\rho_6}^2+81\,{\rho_4}^4+32\,\rho_1{\rho_6}^2
\rho_3-24\,{\rho_3}^4\rho_4 \non\\
&
+12\,\rho_1\rho_6{\rho_3}^{3}\Big)
+2048{\beta}^{10}\rho_3{\rho_4}^2\Big( {\rho_3}^2+3\,\rho_6 \Big)
+\frac{4096}{27}\beta^{12}\,{\rho_6}^2\Big( 3\,{\rho_3}^2+8\,\rho_6 \Big)
 \ .
\end{align}
Now the boundary of the configuration space of the trigonometric
$F_4$ model is confined by the algebraic surfaces
(\ref{e5.C})--(\ref{e5.D}) of the fourth and sixth orders,
correspondingly,\footnote{Being in total the algebraic surface of
the tenth order.} unlike the $\tau$-variables where they were of
the third and eighth orders, respectively
(cf.(\ref{e5.C})--(\ref{e5.D})).

\section{Conclusion}

\indent

We have found that the general rational and trigonometric $F_4$
integrable models with two arbitrary coupling constants are
exactly-solvable. After gauging away the ground state
eigenfunction these models look very much alike when written in a
certain Weyl-invariant variables. Their Hamiltonians preserve the
same flag of the spaces of polynomials and both models are
characterized by the same hidden algebra -- each Hamiltonian can
be written as a non-linear combination of generators. It is an
infinite-dimensional but finite-generated Lie algebra of the
differential operators, which we call $f^{(4)}$ algebra. It is
quite interesting that $D_4$ rational and trigonometric models
possess {\em two} hidden algebras: $gl(5)$ and $f^{(4)}$. Similar
situation takes place for $A_2$, $BC_2$ rational and trigonometric
models as well as for $G_2$ rational model. Their hidden algebras
were $gl(3)$ and $g^{(2)}$, respectively, however, $G_2$
trigonometric model had the only hidden algebra $g^{(2)}$ (see
\cite{Turbiner:1998}).

Present work complements previous studies where the algebraic and the
Lie-algebraic forms as well as the corresponding flags were found for
the rational and trigonometric Olshanetsky-Perelomov Hamiltonians
of $ABCD$ series (and their supersymmetric generalizations)
\cite{Ruhl:1995, Brink:1997} and $G_2$ model \cite{Rosenbaum:1998}.
In order to conclude a study of the whole set of
Olshanetsky-Perelomov integrable systems appearing in the Hamiltonian
reduction method it is necessary to
perform the same analysis for remaining $E_{6,7,8}$ integrable rational
and trigonometric models. We consider it as a challenging task for future.

Our concrete consideration does not confirm some results presented
in \cite{Ruehl:1998, Ruehl:1999} for the rational and
trigonometric $F_4$ models. In particular, we found that the
minimal flags of spaces of polynomials preserved by the rational
and trigonometric $F_4$ Hamiltonians coincide, however, for both
cases they differ from those given in \cite{Ruehl:1998,
Ruehl:1999}. For trigonometric case our Weyl-invariant variables
leading to pure triangular form of the $F_4$ Hamiltonian are
singular in $\bet$. It reflects the fact that the triangular form
for rational and trigonometric models has a different origin: for
the rational case the diagonal matrix elements are proportional to
$\om$ while for trigonometric one to $\bet$. The variables
(\ref{e5.9}) look different from those presented in a general
scheme \cite{Ruehl:1999, Khastgir:00} formulated for all root
systems.

\small{
\section*{Acknowledgement}
J.C.L. and A.V.T. express their gratitude to M.~Rosenbaum for the
extensive discussions on early stage of the work (1997-1998).
A.V.T. thanks N.~Nekrasov for useful conversations and interest to
the work. K.G.B. thanks ICN (UNAM) for kind hospitality where the
collaboration was initiated, K.G.B. and J.C.L. also thanks LPT
(Universit\'e Paris-Sud) where the collaboration was continued,
A.V.T. thanks ITAMP (Harvard University) for inspiring interaction
with its members where the present work was almost finished and
NSF grant for ITAMP at Harvard University for a partial support.
The work was also supported in part by DGAPA grant No. {\it
IN120199}, CONACyT grant {\it 25427-E}, INTAS grant {\it
00-00366}, NATO grant {\it PSTCLG 977275}, and by Russian Fund of
Basic Research 00-15-96786 and 01-02-17383. }
\newpage

\appendix
\setcounter{equation}{0}

\section{Representation of the algebra $gl(5)$}

The algebra $gl(5)$ has a realization in terms of first order
differential operators in four-dimensional $( s_1, s_2, s_3,
s_4)$-space:
\begin{align}\label{ea1.1}
 J_i^- &= \pa_i\ ,\quad i=1,2,3,4 \ ,
\non\\[2pt]
 J_{ik}^0 &=  s_i\pa_k\ ,\quad i,k=1,2,3,4\ ,
\non\\[2pt]
 J^0 &= n - \sum  s_i\pa_i \ ,
\non\\[2pt]
 J_i^+ &=  s_i J^0 \ ,\quad i=1,2,3,4 \ ,
\end{align}
where $n\in C$ is a free parameter. In this realization a grading
$(a_i|\,i=1,2,3,4)$ can be assigned to the generators (A.1)
through their action on monomial
\[
J  s_1^{p_1}  s_2^{p_2}  s_3^{p_3}  s_4^{p_4} \propto  s_1^{p_1+a_1}
 s_2^{p_2+a_2}  s_3^{p_3+a_3}  s_4^{p_4+a_4}\ .
\]
Then the grading of $J$ is defined as a four-component vector
$\vec{a}=(a_1,a_2,a_3,a_4)$.

If $n$ is a non-negative integer, the representation (A.1) becomes
finite-dimensional and the corresponding representation space is a
linear space of polynomials
\begin{equation}
\label{ea1.2}
 {\cal P}_n=\langle  s_1^{p_1}  s_2^{p_2}  s_3^{p_3}  s_4^{p_4} |\ 0\leq
(p_1+p_2+p_3+p_4) \leq n \rangle\ .
\end{equation}
For fixed $n$ the algebra (A.1) acts on the space (A.2)
irreducibly. As a function of $n$ the spaces ${\cal P}_n$ possess
a property that ${\cal P}_n \subset {\cal P}_{n+1}$ for each $n
\in \mathbb{Z}_+$ and form an infinite flag (filtration)
$$\bigcup_{n \in \mathbb{Z}_+} {\cal P}_n = \mathcal{P}.$$

\setcounter{equation}{0}

\section{The $f^{(4)}$ algebra}

We define the algebra $f^{(4)}$ as an algebra of differential
operators on $\mathbb{C}^4$  which acts irreducibly on the space
of inhomogeneous polynomials in four variables
\begin{equation}
\label{b.1}
 {\cal P}_{n} \ = \ \langle  s_1^{p_1}
  s_2^{p_2}  s_3^{p_3}  s_4^{p_4} |\
 0 \leq p_1 + 2 p_2 + 2 p_3 + 3 p_4 \leq n \rangle\ ,
\end{equation}
where $n \in \cal{N}$. Thus, $f^{(4)} \subset
{\mbox{diff}}({\mathbb C}^4)$.

The structure of the algebra $f^{(4)}$ is the following. It
contains three abelian subalgebras $R^{(k)},\ k=2,3,4$ of first
order differential operators
\begin{align}\label{b.2}
 R^{(2)}_i &=  s_1^i\pa_{2}\ ,\ i=0,1,2 \ ,\non\\
 \quad R^{(3)}_i &=  s_1^i\pa_{3}\ ,\ i=0,1,2\ ,
\non\\
  R^{(4)}_i &=  s_1^i\pa_{4}\ ,\ i=0,1,2,3 \ ,
\end{align}
and 11-dimensional subalgebra $B$ of first order differential
operators
\begin{alignat}{4}\label{b.3}
 B^{(1)}_0 &= \pa_{1}\ ,&\qquad  B^{(1)}_1 &=  s_1 \pa_{1}\ ,&\qquad B^{(2)}_2 &=
 s_2\pa_{2}\ ,&\qquad B^{(2)}_3 &=  s_3\pa_{2} \ ,
\non\\
 B^{(3)}_2 &=  s_2\pa_{3}\ ,& B^{(3)}_3 &=  s_3\pa_{3}\ ,
 & B^{(4)}_2 &=  s_2\pa_{4}\ ,& B^{(4)}_3 &=  s_3\pa_{4}\ ,
\non\\
 B^{(4)}_4 &= s_4\pa_{4}\ ,& B^{(4)}_{12} &=  s_1  s_2\pa_{4}\ ,
 & B^{(4)}_{13} &=  s_1  s_3\pa_{4}\ . &
\end{alignat}
These algebras form a subalgebra $B \ltimes (R^{(2)} \oplus
R^{(3)}\oplus R^{(4)}) \subset f^{(4)}$. Also there exists a set
of the second order differential operators
\begin{alignat}{4}
T^{(11)}_2 &=  s_2\pa^2_{11}\ ,&\quad T^{(11)}_3 &=  s_3\pa^2_{11}\ ,
&\quad T^{(12)}_4 &=  s_4\pa^2_{12}\ ,&\quad T^{(13)}_4 &=  s_4\pa^2_{13}\ ,
\non\\
T^{(22)}_4 &=  s_4\pa^2_{22}\ ,& T^{(22)}_{14} &=  s_1  s_4\pa^2_{22}\ ,
& T^{(23)}_4 &=  s_4\pa^2_{23}\ ,& T^{(23)}_{14} &=  s_1  s_4\pa^2_{23}\ ,
\non\\
T^{(33)}_4 &=  s_4\pa^2_{33}\ ,& T^{(33)}_{14} &=  s_1  s_4\pa^2_{33} \ ,
& T^{(14)}_{22} &=  s_2^2 \pa^2_{14}\ ,& T^{(14)}_{23} &=  s_2 s_3\pa^2_{14}\ ,
\non\\
T^{(14)}_{33} &=  s_3^2 \pa^2_{14}\ ,& T^{(44)}_{222} &=  s_2^3 \pa^2_{44}\ ,
& T^{(44)}_{223} &=  s_2^2 s_3\pa^2_{44}\ ,
& T^{(44)}_{233} &=  s_2  s_3^2 \pa^2_{44}\ ,
\non\\
 T^{(44)}_{333} &=  s_3^3 \pa^2_{44}\ ,
\end{alignat}
and third order differential operators
\begin{alignat}{4}
T^{(111)}_4 &=  s_4\pa^3_{111}\ ,&\quad T^{(222)}_{44} &=  s_4^2\pa^3_{222}
\ ,\quad T^{(223)}_{44} &=  s_4^2\pa^3_{223}\ ,&\quad
T^{(233)}_{44} &=  s_4^2\pa^3_{233} \ ,
\non\\
\label{b.4}
 T^{(333)}_{44} &=  s_4^2\pa^3_{333}\ .
\end{alignat}
One can show that the infinite-dimensional algebra generated by 43
generators $R$, $B$, $T$ possesses infinitely many common
invariant subspaces: it leaves invariant the space ${\cal P}_{n}$ for
any $n \in \cal{N}$ and therefore preserves the flag (\ref{e4.9}).

Let us introduce an auxiliary generator
\[
J^0 =  s_1 \pa_1 + 2  s_2 \pa_2 + 2  s_3 \pa_3 + 3  s_4 \pa_4 - n\ .
\]
Then one can define six `raising' generators
\begin{alignat}{3}
\label{b.5}
 J^+_1 &=  s_1 J^0\ ,&\qquad J^+_{3,-2} &=  s_3 \pa_2 J^0\ ,
 &\qquad J^+_{4,-2} &=  s_4 \pa_2 J^0\ ,
\non\\
  J^+_{22,-4} &=  s_2^2 \pa_4 J^0\ ,
 & J^+_{23,-4} &=  s_2  s_3 \pa_4 J^0\ ,
 & J^+_{22,-2} &=  s_2^2 \pa_4 J^0\ .
\end{alignat}
Finally, we get the infinite-dimensional algebra generated by 49
generators (\ref{b.2}) -- (\ref{b.5}) and it is called by
definition $f^{(4)}$.

These raising generators (\ref{b.5}) determine a highest-weight
vector if $n$ is a non-negative integer number. It leads to a
finite-dimensional representation. It can be demonstrated that the
operators (\ref{b.2}) -- (\ref{b.5}) leave the space (\ref{b.1})
invariant at fixed $n$ and act on it irreducibly. Hence, as stated
by the Burnside theorem \cite{lang}, any operator acting on
(\ref{b.1}) allows a representation as a non-linear combination of
the operators (\ref{b.2}) -- (\ref{b.5}) plus an operator
annihilating (\ref{b.1}) (annihilator). Therefore, the
endomorphism of the space (\ref{b.1}) is given by the
infinite-dimensional algebra  $f^{(4)}$ generated by 49 generators
(\ref{b.2}) -- (\ref{b.5}). In turn,  subalgebra of $f^{(4)}$
generated by the generators (\ref{b.2}) -- (\ref{b.4}) possesses
infinitely many finite-dimensional invariant subspaces (\ref{b.1})
at $n=0,1,\ldots$ and hence preserve the infinite non-classical
flag of spaces of polynomials (\ref{e4.9}), $ {\cal P}_0 \subset
{\cal P}_1 \subset {\cal P}_2 \subset \dots \subset {\cal P}_n
\subset \dots\ $.

\section{Dual relations}

In this Appendix we present polynomial functions $g_a(\check{\si};\beta)$
which satisfy the eq.(\ref{dual}) at
$\check{\si}_k = S_k(\sin^2 (\beta z_k/2)/\beta^2)$.
These functions can be used as variables providing an algebraic form
for the $F_4$ Hamiltonian in the $z$-representation (\ref{e3.2}).
Compare these formulas with much simpler form of the functions
$\tau_a = f_a(\si;\beta)$ from eq.(\ref{e5.4}) which give
an algebraic form to the Hamiltonian in $x$-representation.
\begin{align}
 g_1(\check{\si};\beta) &=\, 2\check{\si}_1-
 \frac{2}{3}\beta^2(2\check{\si}_2+\check{\si}_1^2)+
 \frac{16}{3}\beta^4\check{\si}_3-\frac{32}{3}\bet^6\check{\si}_4 \ ,\non\\[5pt]
 g_3(\check{\si};\beta) &= -8\check{\si}_3+\frac{4}{3}\check{\si}_1\check{\si}_2-
 \frac{1}{3}\check{\si}_1^3
+\frac{1}{18}\beta^2\big(4\check{\si}_2\check{\si}_1^2+\check{\si}_1^4-
 32\check{\si}_2^2+120\check{\si}_1\check{\si}_3\big) \non\\[5pt]
&-\frac{8}{9}\beta^4\big(
6\check{\si}_4\check{\si}_1+2\check{\si}_3\check{\si}_2
+\check{\si}_1^2\check{\si}_3 \big)
+\frac{16}{9}
\beta^6\big(2\check{\si}_4\check{\si}_2+\check{\si}_1^2\check{\si}_4
+2\check{\si}_3^2 \big) \ \non\\[5pt] &-\frac{128}{9}\beta^8
\check{\si}_4\check{\si}_3 +\frac{128}{9}\beta^{10}\check{\si}_4^2
 \ ,\non\\
 g_4(\check{\si};\beta) &= 16\check{\si}_4-\frac{2}{3}\check{\si}_1^2\check{\si}_2
+\frac{4}{3}\check{\si}_2^2+\frac{1}{12}\check{\si}_1^4\
+\frac{2}{3}\beta^2\big(\check{\si}_1^2\check{\si}_3-24\check{\si}_1\check{\si}_4
-4\check{\si}_3\check{\si}_2\big) \non\\[5pt]
&+\frac{4}{3}
\beta^4\big(-\check{\si}_1^2\check{\si}_4+16\check{\si}_4\check{\si}_2
+\check{\si}_3^2\big) -\frac{64}{3}\beta^6
\check{\si}_4\check{\si}_3 +\frac{64}{3}\beta^8\check{\si}_4^2 \
,\non\\[5pt]
 g_6(\check{\si};\beta) &= \frac{16}{9}\check{\si}_2^3-\frac{4}{3}\check{\si}_1^2\check{\si}_2^2
-\frac{1}{36}\check{\si}_1^6+16\check{\si}_1^2\check{\si}_4-64\check{\si}_4\check{\si}_2
+\frac{1}{3} \check{\si}_2\check{\si}_1^4 \non\\[5pt]
&+\frac{1}{3}\bet^2\big(192\check{\si}_4\check{\si}_3+192\check{\si}_4\check{\si}_2\check{\si}_1
-48\check{\si}_4\check{\si}_1^3-16\check{\si}_3\check{\si}_2^2
+8\check{\si}_3\check{\si}_2\check{\si}_1^2-\check{\si}_3\check{\si}_1^4
\big) \non\\[5pt]
&+\frac{2}{3}\beta^4\big(-192\check{\si}_4^2-96\check{\si}_4\check{\si}_3\check{\si}_1
-80\check{\si}_4\check{\si}_2^2+16\check{\si}_4\check{\si}_2\check{\si}_1^2
+\check{\si}_4\check{\si}_1^4+8\check{\si}_3^2\check{\si}_2
\non\\[5pt] &-2\check{\si}_3^2\check{\si}_1^2\big)
 +\frac{16}{9}
\beta^6\big(72\check{\si}_4^2\check{\si}_1+60\check{\si}_4\check{\si}_3\check{\si}_2
-6\check{\si}_4\check{\si}_3\check{\si}_1^2-\check{\si}_3^3 \big)
\non\\[5pt]
&+\frac{32}{3}\bet^8\big(-16\check{\si}_4^2\check{\si}_2+\check{\si}_4^2\check{\si}_1^2
-5\check{\si}_4\check{\si}_3^2\big) +\frac{512}{3}\beta^10
\check{\si}_4^2\check{\si}_3 -\frac{1024}{9}\beta^12 \check{\si}_4^3
 \ .
\end{align}

\section{First eigenfunctions of the rational and
trigonometric $F_4$ models}

In this Appendix we present explicit expressions for the first
eigenfunctions of $F_4$ models at $n=0,1,2$.

\vskip .4cm

{\bf I.\ \it Rational $F_4$ model.}
\nopagebreak
\noindent
\begin{alignat*}{3}
 &\bullet\ {\bf n=0} &\\
 &\qquad \Phi_0 =&& 1\ ,\\
 &\qquad E_0 =&& 0 \ .  \\[8pt]
 &\bullet\ {\bf n=1} &\\
 &\qquad \Phi_1 =&&\  s_1 - \frac{2}{\om}(6\mu+6\nu+1)\ , \\
 &\qquad E_1 =&& -4\om \ . \\[8pt]
 &\bullet\ {\bf n=2} \\
 &\qquad \Phi_2^{(1)} =&&\  s_1^2\ -\ \frac{6}{\om}(4\mu+4\nu+1) s_1\ +\
 \frac{6}{\om^2}(4\mu+4\nu+1)(6\mu+6\nu+1)\ , \\
 &\qquad E_2^{(1)} =&& - 8\om\ , \\[5pt]
 &\qquad \Phi_2^{(2)} =&&  s_3\ +\ \frac{1}{4\om}(2\mu+4\nu+1) s_1^2\ -\
\frac{3}{4\om^2}(2\mu+4\nu+1)(4\mu+4\nu+1) s_1\ \\
&&&
 + \frac{1}{2\om^3}(2\mu+4\nu+1)(6\mu+6\nu+1)(4\mu+4\nu+1)\ , \\
&\qquad E_2^{(2)} =&& -12\om\ ,\\[5pt]
&\qquad
\Phi_2^{(3)} =&&\  s_4 + \frac{1}{\om}(3\nu+1) s_3\ +\
\frac{1}{8\om^2}(3\nu+1)(2\mu+4\nu+1) s_1^2\ \\
&&&
 - \frac{1}{4\om^3}(3\nu+1)(2\mu+4\nu+1)(4\mu+4\nu+1) s_1\  \\
&&&
 +\frac{1}{8\om^4}(3\nu+1)(2\mu+4\nu+1)(6\mu+6\nu+1)(4\mu+4\nu+1)\ ,\\
 &\qquad E_2^{(3)} =&& -16\om
\end{alignat*}


\vskip .4cm

{\bf II.\ \it Trigonometric $F_4$ model.}
\nopagebreak
\noindent
\begin{alignat*}{3}
 &\bullet\ {\bf n=0} &\\
 &\qquad \Phi_0 =&& 1\ ,\\
 &\qquad E_0 =&& 0 \ .  \\[8pt]
 &\bullet\ {\bf n=1} &\\
 &\qquad \Phi_1 =&&\ \rho_1 - \frac{6\nu + 6\mu +1}{(5\nu + 6\mu +1)}\bet^{-2} \ ,\\
 &\qquad E_1 =&& 4(5\nu  + 6\mu  + 1)\bet^2 \ .  \\[8pt]
 &\bullet\ {\bf n=2} \\
 &\qquad \Phi_2^{(1)} =&&\ \rho_3 +
\frac{3}{8}\frac{(4\nu+4\,\mu+1)}{(\nu +4\mu +1)}\bet^{-4}\rho_1 -
\frac{3}{16}\frac{(4\nu+4\mu+1)(6\nu +6\mu +1)}{(\nu+4\mu
+1)(3\nu+5\mu +1)}\bet^{-6} \ ,\\
 &\qquad E_2^{(1)} =&& 8\,(3\,\nu  + 5\,\mu  + 1)\bet^2 \ .\\[5pt]
 &\qquad \Phi_2^{(2)} =&& \frac{2}{3}(3\nu + 2 \mu  + 1)\bet^2 \rho_4 + \nu\rho_3 +
\frac{3}{16}\frac {(4\nu + 4\mu  + 1)(4\nu + 2\mu  + 1)}{(2\nu
+3\mu  + 1)}\bet^{-4}\rho_1  \\ &&& \mbox{} -
\frac{1}{16}\frac{(4\nu +4\mu +1)(4 \nu+2\mu + 1)(6\nu  + 6\mu +
1)}{(2\nu + 3\mu + 1)(3\nu  + 4\,\mu  + 1)}\bet^{-6} \ ,\\ &\qquad
E_2^{(2)} =&& 12(3\nu  + 4\mu  + 1)\bet^2 \ .\\[5pt]
 &\qquad \Phi_2^{(3)} =&&\ \rho_4 + \frac {3}{8} \frac {(3\,\nu  +
1)}{(2\,\nu  + \mu + 1)}\bet^{-2}\rho_3  + \frac {9}{32} {(\nu
+1)}\bet^{-4}\rho_1^2  \\ &&& \mbox{} - \frac{9}{64}\frac {(3\nu
+1)(4\nu+4\mu+1)(4\nu+2\mu+1)} {(5\nu+6\mu+3)(2\nu +\mu
+1)}\bet^{-6}\rho_1 \\ &&& \mbox{} + \frac{9}{128}\,\frac{(3\,\nu  +
1)\,(4\,\nu + 4\, \mu  + 1)\,(4\,\nu  + 2\,\mu  + 1)\,(6\,\nu  +
6\,\mu  + 1)}{(5\, \nu  + 6\,\mu  + 3)\,(2\,\nu  + \mu  +
1)\,(5\,\nu  + 6\,\mu  + 2 )}\bet^{-8} \ , \\
 &\qquad E_2^{(3)} =&& 8(5\nu  + 6\mu  + 2)\bet^2 \ .
\end{alignat*}

\newpage

\begingroup\raggedright
\endgroup

\end{document}